\documentclass[aip,amsmath,amssymb,reprint,nobalancelastpage,floatfix]{revtex4-2}

\usepackage{graphicx}
\usepackage{dcolumn}
\usepackage{bm}

\usepackage[utf8]{inputenc}
\usepackage[T1]{fontenc}
\usepackage{etoolbox}
\usepackage{orcidlink}
\usepackage{lipsum}
\usepackage{physics}
\usepackage[caption=false]{subfig}
\newcommand{\phantomsubfloat}[1]{{\captionsetup[subfigure]{labelformat=empty}\subfloat{#1}}}
\usepackage{tikz}
\usepackage{svg}
\usepackage{layouts}

\makeatother

\newcommand{\hide}[1]{}

\newcommand{\oqupy}{{OQuPy}}
\newcommand{\code}[1]{\texttt{#1}}
\newcommand{\citeTEMPOmethods}{\cite{Strathearn2017, Strathearn2018, Gribben2021,  Chiu2021}}
\newcommand{\citePTmethods}{\cite{ Fux2021, Gribben2022, fux_tensor_2023, fowler2022, butler_optimizing_2024}}
\newcommand{\citeallmethods}{\cite{Strathearn2017, Strathearn2018, Gribben2021, Chiu2021, Fux2021, Gribben2022, fux_tensor_2023, fowler2022, butler_optimizing_2024}}
\newcommand{\citePTcreation}{\cite{Banuls2009, muller-hermes_tensor_2012, Strathearn2019, Jorgensen2019, Lerose2020, Sonner2021, Cygorek2021, Ye2021, thoenniss_efficient_2022, Thoenniss2023, ng_real_2022, cygorek2023, link2023, Kloss2023, Ng2023}}

\newcommand{\ps}{\ensuremath{\mathrm{ps}}}
\newcommand{\ips}{\ensuremath{\mathrm{ps}^{-1}}}

\begin{document}

\preprint{AIP/123-QED}

\title{\oqupy{}: A Python package to efficiently simulate non-Markovian open quantum systems with process tensors}
\author{Gerald E. Fux\,\orcidlink{0000-0002-7912-0501}}
\affiliation{The Abdus Salam International Center for Theoretical Physics (ICTP), Strada Costiera 11, 34151 Trieste, Italy}
\author{Piper Fowler-Wright\,\orcidlink{0000-0003-1060-445X}}
\affiliation{SUPA, School of Physics and Astronomy, University of St Andrews, St Andrews, KY16 9SS}
\author{Joel Beckles\,\orcidlink{0009-0003-5837-307X}}
\affiliation{SUPA, School of Physics and Astronomy, University of St Andrews, St Andrews, KY16 9SS}
\author{Eoin P. Butler\,\orcidlink{0000-0002-3888-5479}}
\affiliation{School of Physics, Trinity College Dublin, Dublin 2, Ireland}
\affiliation{Trinity Quantum Alliance, Unit 16, Trinity Technology and Enterprise Centre, Pearse Street, Dublin 2, Ireland}
\author{Paul R. Eastham\,\orcidlink{0000-0002-7054-1457}}
\affiliation{School of Physics, Trinity College Dublin, Dublin 2, Ireland}
\affiliation{Trinity Quantum Alliance, Unit 16, Trinity Technology and Enterprise Centre, Pearse Street, Dublin 2, Ireland}
\author{Dominic Gribben\,\orcidlink{0000-0003-0649-1552}}
\affiliation{Institute for Physics, Johannes Gutenberg University of Mainz, D-55099 Mainz, Germany}
\author{Jonathan Keeling\,\orcidlink{0000-0002-4283-552X}}
\affiliation{SUPA, School of Physics and Astronomy, University of St Andrews, St Andrews, KY16 9SS}
\author{Dainius Kilda\,\orcidlink{0000-0002-1918-3172}}
\affiliation{Max-Planck-Institut für Quantenoptik, Hans-Kopfermann-Strasse 1, D-85748 Garching, Germany}
\author{Peter Kirton\, \orcidlink{0000-0002-3915-1098}}
\affiliation{Department of Physics and SUPA, University of Strathclyde, Glasgow, G4 0NG, United Kingdom}
\author{Ewen D.C. Lawrence\, \orcidlink{0009-0004-6528-016X}}
\affiliation{Department of Physics and SUPA, University of Strathclyde, Glasgow, G4 0NG, United Kingdom}
\author{Brendon W. Lovett\,\orcidlink{0000-0001-5142-9585}}
\affiliation{SUPA, School of Physics and Astronomy, University of St Andrews, St Andrews, KY16 9SS}
\author{Eoin O'Neill\,\orcidlink{0000-0003-2915-0689}}
\affiliation{School of Physics, Trinity College Dublin, Dublin 2, Ireland}
\affiliation{Trinity Quantum Alliance, Unit 16, Trinity Technology and Enterprise Centre, Pearse Street, Dublin 2, Ireland}
\author{Aidan Strathearn}
\affiliation{Institute for Photonics and Advanced Sensing (IPAS) and School of Physical Sciences,
University of Adelaide, Adelaide, South Australia 5005, Australia}
\author{Roosmarijn de Wit\,\orcidlink{0009-0000-1858-0064}}
\affiliation{SUPA, School of Physics and Astronomy, University of St Andrews, St Andrews, KY16 9SS}

\date{\today}

\begin{abstract}
Non-Markovian dynamics arising from the strong coupling of a system to a structured environment is essential in many applications of quantum mechanics and emerging technologies.
Deriving an accurate description of general quantum dynamics including memory effects is however a demanding task, prohibitive to standard analytical or direct numerical approaches.
We present a major release of our open source software package, \oqupy{} (Open Quantum System in Python), which provides several recently developed numerical methods that address this challenging task.
It utilizes the process tensor approach to open quantum systems in which a single map, the process tensor, captures all possible effects of an environment on the system.
The representation of the process tensor in a tensor network form allows an exact yet highly efficient description of non-Markovian open quantum systems (NM-OQS).
The \oqupy{} package provides methods to (1) compute the dynamics and multi-time correlations of quantum systems coupled to single and multiple environments, (2) optimize control protocols for NM-OQS, (3) simulate interacting chains of NM-OQS, and (4) compute the mean-field dynamics of an ensemble of NM-OQS coupled to a common central system.
Our aim is to provide an easily accessible and extensible tool for researchers of open quantum systems in fields such as quantum chemistry, quantum sensing, and quantum information.
\end{abstract}

\maketitle


\section{Introduction}
The theory of open quantum systems is one of the key ingredients for making quantum mechanics more applicable for experimental applications.
It describes the dynamics of microscopic systems governed by the rules of quantum mechanics, but unlike the theory of closed systems, open systems additionally incorporate the unavoidable interactions with the environment.
Many physical scenarios allow for simple time-local effective equations of motion~\cite{Gardiner2000, Breuer2002}.
Such a description is called \emph{Markovian}, which means that the future evolution of the system only depends on its present state, and not on its history.
This is typically a good approximate description when the environment is featureless or the interaction between the system and environment is weak.
However, there are numerous applications where such a Markovian description fails.
Solid state based quantum devices, such as quantum dots and nitrogen-vacancy color centers in diamonds, for example, often interact strongly with the lattice vibrational modes, surrounding charges, or magnetic spins~\cite{Chirolli2008}.
In many such cases a non-Markovian treatment is necessary~\cite{WilsonRae2002, Galland2008, Chirolli2008, ramsay2010, Roy2011, DeVega2017}.
Non-Markovian descriptions of open quantum systems are also an important ingredient for the development of better light-harvesting devices~\cite{Meng2018} and understanding the role of quantum mechanics in biological systems~\cite{cao2020}, such as the Fenna-Matthews-Olson complex~\cite{Engel2007, wang2019}.
Apart from these particular applications, the general theory of non-Markovian open quantum systems is also of importance for fundamental research, such as the study of strong-coupling quantum thermodynamics~\cite{Nicolin2011, Horodecki2013, Vinjanampathy2016, Seifert2016, Mitchison2019, Binder2019, Talkner2020} and the development of theoretical tools for the characterization of quantum devices~\cite{White2021}.

\begin{figure*}
	\includegraphics[width=0.98\textwidth]{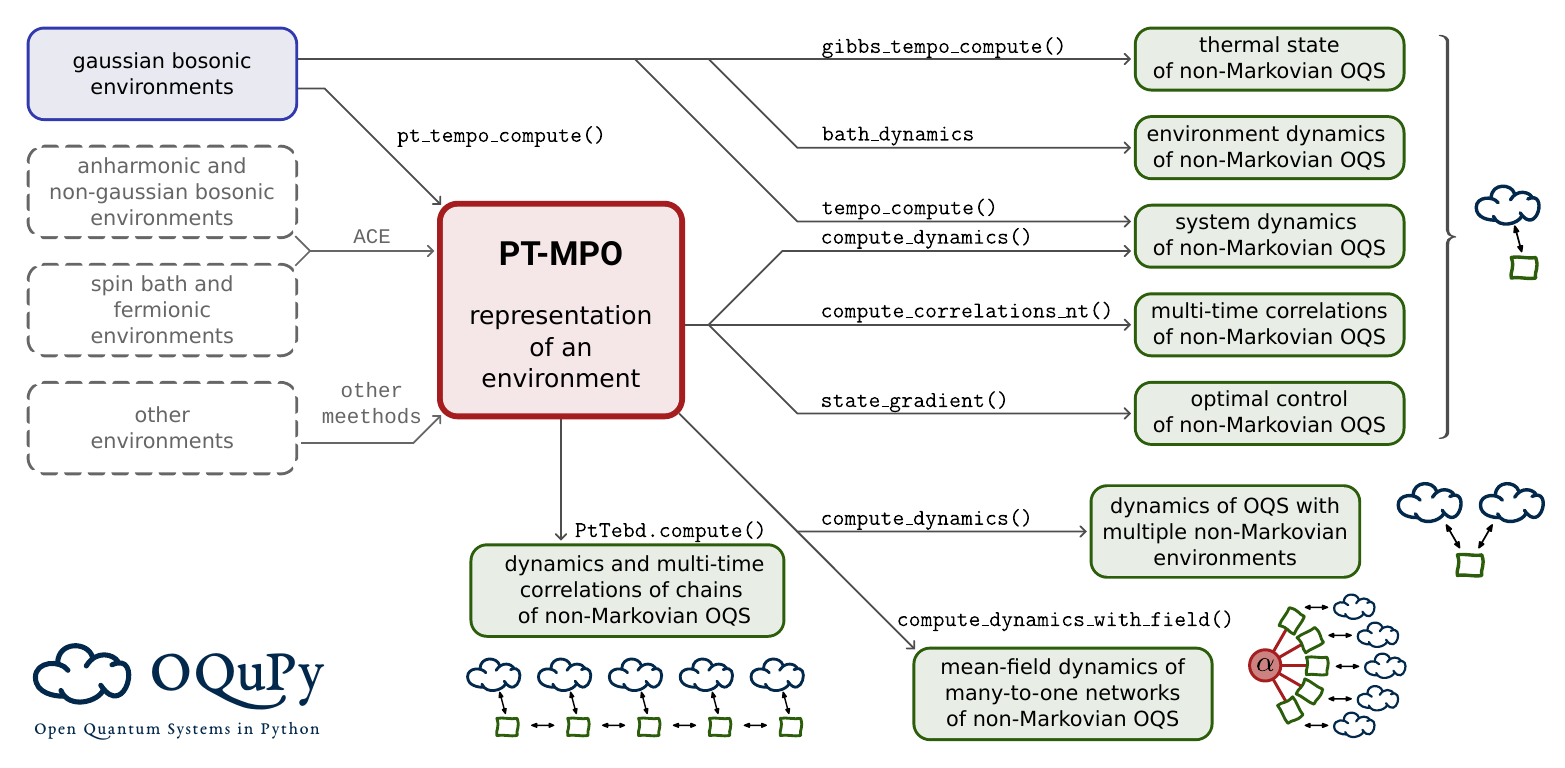}
	\caption{\label{fig:computation-flow}%
        Schematic of the workflow for various types of use cases in \oqupy{}.
        Each green box corresponds to a particular type of scenario of interest, most of which require a PT-MPO representation of the relevant environment (red box).
        The construction of PT-MPOs is available in \oqupy{} for Gaussian bosonic environments (blue box).
        One can import PT-MPOs created with other methods external to \oqupy{} (gray dashed boxes), such as the automated compression of environments (ACE) method~\cite{Cygorek2021}.
        For each type of use case scenario we display a sketch of the interactions among the open quantum systems (green squares), the environments (blue clouds), and possibly a common central system (red circle).
    }
\end{figure*}

In recent years much progress has been made in the development of numerical methods to treat non-Markovian open quantum systems.%
\vphantom{\citeallmethods{}} 
A particularly versatile class of recently developed methods~\citePTmethods{} is based on an operational approach to open quantum systems, known as the \emph{process tensor framework}~\cite{Pollock2018, milz_quantum_2021}, which is well suited for the challenges of non-Markovianity.
Related to this approach are several other recently developed tensor network methods\citeTEMPOmethods{} that build on the Quasi Adiabatic Path Integral (QUAPI)\cite{Makri1995, Makri1995a, kundu_pathsum_2023}.

The purpose of the \oqupy{} \code{Python} package is to give easy access to several of these recently developed methods~\citeallmethods{} that combine the conceptional advantages of the process tensor framework, with the numerical efficiency of tensor networks.
In this article we give an overview of the toolset available in \oqupy{}.
This includes numerically exact methods (i.e.~employing only numerically well controlled approximations) for the non-Markovian dynamics and multi-time correlations of
\begin{itemize}
	\setlength\itemsep{-1mm}
	\item quantum systems coupled to a single environment~\cite{Strathearn2017, Strathearn2018, Jorgensen2019},
	\item quantum systems coupled to multiple environments~\cite{Gribben2022},
	\item interacting chains of non-Markovian open quantum systems~\cite{fux_tensor_2023}, and
	\item ensembles of open many-body systems with many-to-one coupling to some common central system~\cite{fowler2022}.
\end{itemize}
Furthermore, \oqupy{} implements methods to
\begin{itemize}
	\setlength\itemsep{-1mm}
	\item optimize control protocols for non-Markovian open quantum systems~\cite{Fux2021, butler_optimizing_2024},
	\item compute the dynamics of an non-Markovian environment~\cite{Gribben2021}, and
	\item obtain the thermal state of a quantum system strongly coupled to a structured environment~\cite{Chiu2021}.
\end{itemize}

At the heart of most of these methods lies the numerically efficient representation of the involved environment through a so-called \emph{Process Tensor in Matrix Product Operator form} (PT-MPO), which we will introduce in Section~\ref{sec:process-tensors} below.
In short, the PT-MPO representation of an environment allows the systematical removal of negligible correlations between the system and environment, leading to a numerically efficient characterization of their interaction.
The overall workflow for most of the methods in \oqupy{} is thus separated into two stages:
first, the computation of the PT-MPO representation for the environment of interest, and second, the application of the PT-MPO in one of the scenarios listed above.
We sketch this workflow diagrammatically in Fig~\ref{fig:computation-flow}.

\textit{Outline.---}%
In this article we give an overview of the underlying methodology and practical application of the \oqupy{} package. 
We begin in Section~\ref{sec:process-tensors} by introducing the process tensor approach to open quantum systems.
In particular, we present the construction of PT-MPOs using a \emph{process tensor adoption} of the so-called \emph{time evolving matrix product operator} method (PT-TEMPO) and comment on its computational parameters in Section~\ref{sub:pt-tempo}. 
In Section~\ref{sec:use-cases} we give a brief introduction to the various methods built around PT-MPOs and showcase their application in the \oqupy{} package.
This includes methods for the dynamics of non-Markovian open quantum systems in Section~\ref{sub:simple-dynamics}, multi-time correlations of non-Markovian open quantum systems in Section~\ref{sub:multi-time-correlations}, optimization of control protocols in Section~\ref{sub:optimization}, and the dynamics of open quantum systems with multiple environments in Section~\ref{sub:multi-environments}.
Also, we briefly introduce methods to address many-body open quantum systems such as chains of open quantum systems in Section~\ref{sub:pt-tebd}, and many-to-one networks of open quantum systems in Section.~\ref{sub:mean-field}.
Section~\ref{sec:other-methods} discusses other related methods that are also included in \oqupy{}, such as a method for the computation of the environment dynamics (Section~\ref{sub:environment-dynamics}), an alternative method for the system dynamics (Section~\ref{sub:tempo}), and a method to directly compute the thermal state of a non-Markovian open quantum system (Section~\ref{sub:gibbs-tempo}).
Finally we conclude with a brief discussion on possible future directions for further development of the \oqupy{} package in Section~\ref{sec:future-directions}.

\section{Process Tensor Framework}
\label{sec:process-tensors}
The process tensor framework introduced by Pollock \textit{et al.}~\cite{Pollock2018} is an operational approach to fully characterize general open quantum systems and differs fundamentally from the canonical approach based on dynamical maps and their master equations.
The dynamical map describes the evolution of the reduced system state for each initial state and hence, from an operational point of view, corresponds to prepare--evolve--measure type experiments.
Given that an experiment can only access the system, but has no direct control over the environment, one may be led to believe that such a collection of dynamical maps gives a full description of the operationally accessible quantities of the open quantum system.
This is, however, not true.
Not all properties of an open system can be determined in a prepare--evolve--measure type experiment.
In many experiments the observed quantities are related to multi-time correlations, or, one is interested in measuring the state of the system after multiple interactions with additional ancillary systems.
In fact, many scenarios of interest, such as the optimization of control protocols of an open quantum system, or the interaction among multiple open quantum systems, necessarily require the correct treatment of multi-time correlations.
However, the dynamical map---even if known exactly from an exact master equation---cannot encode the necessary information to correctly describe multi-time correlations~\cite{milz_quantum_2021}.

The process tensor, on the other hand, not only encodes the outcome of prepare--evolve--measure type experiments, but also encodes the outcome of all possible experiments with intermediate interventions on the system, and thus also multi-time correlations.
It is closely related to quantum combs~\cite{Chiribella2008}, generalized influence functionals~\cite{Jorgensen2019}, and process matrices~\cite{Oreshkov2012}.
The process tensor stands in a one-to-one relation to the outcome of all possible experiments one can perform on an open quantum system and is hence an operationally well defined and complete characterization of general open quantum systems.
While the dynamical map is the map from all possible initial system states to the resulting final system state, the process tensor is the map from all possible control sequences to the final system state.
Consider, for example, an open quantum system where one prepares the system in a certain state at time $t_0$, applies a projective measurement in the $X$ basis at time $t_1$, and another projective measurement in the $Z$ basis at time $t_2$.
In such a case the dynamical map correctly encodes the expectation value of the first measurement (i.e. the $X$ measurement).
However, the expectation value of the $Z$ measurement, as well as the correlations between the two measurement outcomes will generally not be correctly encoded in the dynamical map, even if the dynamical map is known exactly.
The process tensor, on the other hand, is the map that gives the state of the system at the final time $t_2$ for any sequence of control operations at time $t_0$ and $t_1$, which in the above example are the state preparation and the $X$ measurement.

Let us now define control operations and the process tensor a little more formally~(see Ref.~\onlinecite{Pollock2018} for a fully formal definition). 
A control operation in this context is any physical operation on the system, such as the preparation of the system state, a measurement of the system, the application of a unitary, or the application of a unitary on the system and some ancilla system.
Any such physical control operation can be described by a superoperator (i.e. a linear operator that acts on density matrices) that has the mathematical property of complete positivity (called a CP-map)~\cite{Nielsen2002}.
A control sequence is hence a sequence of CP-maps $\{\mathcal{A}_n\}_{n\in 0\ldots N-1}$ applied to the system at times $\{t_n\}_{n \in 0\ldots N-1}$.
Let us write the overall dynamics of the system and environment between time $t$ and $t^\prime$ as an overall unitary operator $\hat{U}(t^\prime,t)$, and let us consider an overall initial state $\rho_0$, which generally need not be separable between system and environment.
When we include the control sequence $\{\mathcal{A}_n\}_{n\in 0\ldots N-1}$ at the times $\{t_n\}_{n \in 0\ldots N-1}$ then the final reduced density matrix of the system at time $t_N$ is
\begin{equation}
\rho^\mathrm{S}(t_N) = \mathrm{Tr}_\mathrm{E}\left[ \mathcal{U}_{t_{N-1}}^{t_N} \mathcal{A}_{N-1} \, \ldots \, \mathcal{U}_{t_1}^{t_2} \mathcal{A}_1 \, \mathcal{U}_{t_0}^{t_1} \mathcal{A}_0 \rho_0 \right] \mathrm{,}
\end{equation}
where $\mathcal{U}_{t}^{t^\prime}\,\rho\, := \hat{U}(t',t) \,\rho\, \hat{U}^\dagger(t',t)$ is the superoperator of the unitary evolution, and $\mathrm{Tr}_\mathrm{E}$ denotes the partial trace over the environment.
The process tensor $\mathcal{T}_{0:N}$ for times $\{t_n\}_{n \in 0\ldots N}$ is then defined to be the map from such sequences of CP-maps to the corresponding final state of the system, i.e.
\begin{equation}
\rho^\mathrm{S}(t_N) =: \mathcal{T}_{0:N}\left[\mathcal{A}_{N-1}, \ldots, \mathcal{A}_1, \mathcal{A}_0 \right] \mathrm{.}
\end{equation}

Note that the discrete number $N$ of control operations does not imply that the evolution was discrete in time.
It only means that we (for simplicity) consider a discrete number of interventions on the system.
The dynamical map, for example, can be obtained from a process tensor with $N=1$, because dynamical maps only describe the outcome of an experiment at \emph{one} point in time after an initial system preparation (even if that one point in time is variable).

\begin{figure}
	\phantomsubfloat{\label{fig:pt-construction-a}}%
	\phantomsubfloat{\label{fig:pt-construction-b}}%
	\phantomsubfloat{\label{fig:pt-construction-c}}%
	\includegraphics[width=0.49\textwidth]{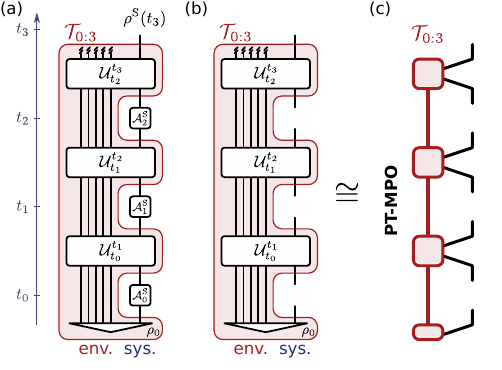}
	\caption{\label{fig:pt-construction}%
        (a)~Quantum circuit for the state at $t_3$ of an open quantum system (sys.) coupled to an environment (env.) with control operations $\mathcal{A}_n$ at times $t_0$, $t_1$, and $t_2$.
        (b)~The process tensor $\mathcal{T}_{0:3}$ of an open quantum system for times $t_0, \ldots, t_3$.
        (c)~The process tensor in matrix product operator form (PT-MPO).
    }
\end{figure}
	
In Fig.~\ref{fig:pt-construction} we show the construction of the process tensor for $N=3$.
It provides the final state of a generic open quantum system at time $t_3$ given control operations at times $t_0$, $t_1$, and $t_2$.
The environment can be of any dimension and is here represented with five lines.
This figure can be read as a quantum circuit or, equivalently, as a tensor network in Liouville space~\cite{Schollwock2011, Orus2014}.
In Fig.~\ref{fig:pt-construction-a} the control sequence $[\mathcal{A}_0, \mathcal{A}_1, \mathcal{A}_2]$ is inserted, yielding the final system state at time $t_3$.
Removing the control sequence and keeping the slots open in Fig.~\ref{fig:pt-construction-b} thus leaves the tensor marked by the red shaded region, which can then be identified with the process tensor $\mathcal{T}_{0:3}$.
Note that the environment Liouville space does not contribute to the dimensionality of the tensor, for which instead only the external legs (corresponding to the system dimension) count. 

\subsubsection*{Process tensor in matrix product operator form (PT-MPO).}
The process tensor is a tensor of dimension $d^{4N+2}$, where $d$ is the Hilbert space dimension of the system.
That is, it grows exponentially with the number of possible interventions $N$.
In many cases, however, it is possible to systematically discard negligible correlations and express the process tensor as a tensor network~\cite{Schollwock2011, Orus2014} in a matrix product operator form~\cite{Verstraete2004, Orus2014} (PT-MPO) as shown in Fig.~\ref{fig:pt-construction-c}.
Here, the single high-rank process tensor is represented as an array of rank-4 tensors connected through indices that are called \emph{bonds}.
The necessary bond dimension of a PT-MPO reflects the degree of non-Markovianity in the interaction~\cite{Pollock2018a}, but does not necessarily scale with the correlation time or the dimension of the environment Hilbert space.
Various methods exist for the construction of PT-MPOs for different types of environments~\citePTcreation{}.
The most straightforward approach is to start with a tensor network that explicitly represents the evolution of the environment and then perform a contraction sequence to yield a PT-MPO~\cite{Banuls2009, muller-hermes_tensor_2012, Sonner2021, Lerose2020}.
For linearly coupled Gaussian environments one can even directly construct a tensor network that yields a PT-MPO without the need to explicitly model the environment as a tensor network~\cite{Jorgensen2019, Strathearn2019, thoenniss_efficient_2022, Thoenniss2023, ng_real_2022, cygorek2023, link2023}.
Other approaches~\cite{Cygorek2021, Ye2021} allow the construction of PT-MPOs for any environment that can be approximated by a finite set of independent degrees of freedom.

All of these methods yield a PT-MPO that fully encodes the influence of an environment on the open system and can then be equally used in various ways (as shown in Fig.~\ref{fig:computation-flow}).
As of version 0.5, the \oqupy{} package implements a single method for the creation of PT-MPOs, namely PT-TEMPO (the process tensor adoption of the time evolving matrix product operator method).
This method allows the computation of PT-MPOs of open quantum systems coupled to Gaussian bosonic environments, such as photon and phonon fields.
We stress, however, that for all applications of PT-MPOs presented in Section~\ref{sec:use-cases} below the origin of the PT-MPO or the method with which it has been created are irrelevant.
PT-MPOs can be created for other types of environments with other methods and other software, and can then be imported into \oqupy{} and applied in the exact same way as PT-MPOs constructed with PT-TEMPO.

\subsection{Construction of PT-MPOs with PT-TEMPO}
\label{sub:pt-tempo}
The time evolving matrix product operator (TEMPO) method is a tensor network method for simulating the reduced dynamics of open quantum systems with Gaussian bosonic environments~\cite{Strathearn2018, Strathearn2019}.
In Ref.~\onlinecite{Jorgensen2019} Jørgensen and Pollock suggest a modification of the TEMPO tensor network contraction scheme to obtain a PT-MPO.
In the following we outline the construction of the TEMPO tensor network and the adopted contraction sequence.

The most general form of the total Hamiltonian we consider for PT-TEMPO is (in units of $\hbar=1$)
\begin{equation}
	\label{eq:tempo-hamiltonian}
	\hat{H}(t) = \hat{H}_\mathrm{S}(t)
	+ \underbrace{\hat{S}\sum_k \left( g_k \hat{b}_k + g_k^* \hat{b}^\dagger_k \right)}_{\hat{H}_\mathrm{I}}
	+ \underbrace{\sum_k \omega_k \hat{b}_k^\dagger \hat{b}_k}_{\hat{H}_\mathrm{E}}  \mathrm{,}
\end{equation}
where $\hat{H}_\mathrm{S}(t)$ is an arbitrary (possibly time dependent) system Hamiltonian, $\hat{S}$ is the system operator that couples to the field, and $\hat{b}_k$ ($\hat{b}_k^\dagger$) are bosonic creation (annihilation) environment operators for the $k^\mathrm{th}$ mode.
Also, we assume that the total initial system and environment state is separable, i.e. $\rho_0 = \rho_0^\mathrm{E} \otimes \rho_0^\mathrm{S}$, where the environment is a Gaussian (e.g. thermal) state with respect to $\hat{H}_\mathrm{E}$.
The environment interaction is fully characterized by the system coupling operator $\hat{S}$ and the spectral density $J(\omega)$, which is defined as
\begin{equation}
	J(\omega)=\sum_k \abs{g_k}^2 \delta(\omega-\omega_k) \mathrm{.}
\end{equation}
In Section~\ref{sec:use-cases}~and~\ref{sec:other-methods} we will often consider $J(\omega)$ to take the power-law form~\cite{Leggett1987}
\begin{equation}
	\label{eq:spectral-density-power-law}
	J(\omega)=\alpha\, \omega^\zeta\, \omega_c^{1-\zeta}\, \exp\left( -\frac{\omega}{\omega_c} \right) \mathrm{,}
\end{equation}
where $\alpha$ is a dimensionless coupling constant, $\zeta$ is the power-law exponent, and $\omega_c$ is a cutoff frequency. For $\zeta=1$ the environment is called ``Ohmic'', for $\zeta<1$ and $\zeta>1$ it is said to be ``sub-Ohmic'' and ``super-Ohmic'', respectively.

\begin{figure*}
	\phantomsubfloat{\label{fig:pt-tempo-network-a}}%
	\phantomsubfloat{\label{fig:pt-tempo-network-b}}%
	\phantomsubfloat{\label{fig:pt-tempo-network-c}}%
	\phantomsubfloat{\label{fig:pt-tempo-network-d}}%
	\includegraphics[width=0.95\textwidth]{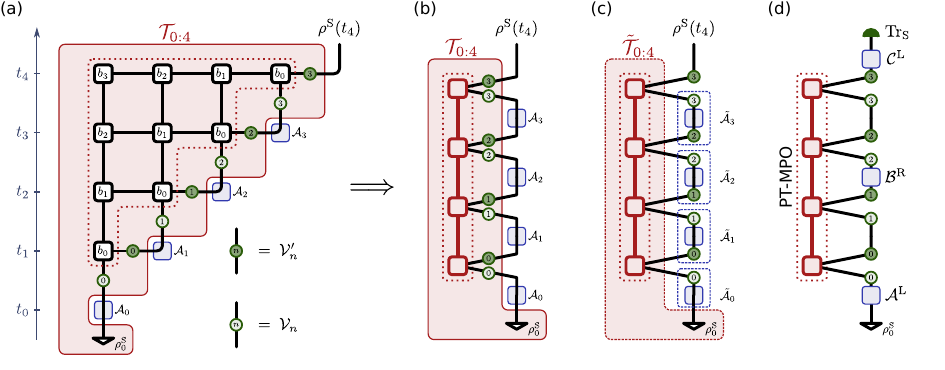}
	\caption{\label{fig:pt-tempo-network}%
        (a)~The TEMPO tensor network for four time steps with influence tensors $b_k$, system propergators $\mathcal{V}^{(\prime)}_n$, a set of control operations $\mathcal{A}_n$, and initial system state $\rho^\mathrm{S}_0$.
        (b)~A PT-MPO $\mathcal{T}_{0:4}$ applied to a set of control operations.
        (c)~A PT-MPO $\tilde{\mathcal{T}}_{0:4}$ applied to a set of control operations and system propagators.
        (d)~A PT-MPO applied to system propagators, three control operations and the trace operation for the computation of three-time correlation $\Tr [\hat{C}(t_4) \hat{A}(t_0) \rho_0 \hat{B}(t_2)]$ as explained in Section~\ref{sub:multi-time-correlations}.
    }
\end{figure*}

The derivation of the TEMPO tensor network~\cite{Strathearn2018} starts with a symmetrized second order Trotter splitting between system and environment interaction of the evolution into $N$ small time steps ($\delta t = t_N / N$).
Tracing out the environment degrees of freedom leads to a discretized version of the Feynman-Vernon influence functional~\cite{Feynman1963}.
The resulting calculation can be represented as a tensor network of the form shown in Fig~\ref{fig:pt-tempo-network-a}.
The tensors $b_k$ are constructed from the coupling operator $\hat{S}$ and the environment's auto-correlation function.
For thermal environments the auto-correlation function is determined by the spectral density $J(\omega)$ and the initial bath temperature $T$ alone.
These tensors quantify how the system evolution at a time step $t_n$ is influenced through the environment by the system's history from time step $t_{n-k}$.
The $\mathcal{V}(n)$ and $\mathcal{V}^\prime(n)$ superoperators are the pure system propagators for the first and second half of the $n^\mathrm{th}$ time step.        
In many cases the two-time auto-correlation function of the environment interaction almost vanishes after some time $\tau_\mathrm{cut}$.
In such a case the influence tensors $b_k$ become identities for $k \delta t > \tau_\mathrm{cut}$ and can be omitted from the tensor network. 

To see how this tensor network links to the process tensor we include additional interventions $\mathcal{A}_n$ to the picture, which we draw as blue squares in Fig.~\ref{fig:pt-tempo-network-a}.
The tensor network in Fig.~\ref{fig:pt-tempo-network-a} now represents the final state $\rho^\mathrm{S}(t_N)$ for any intervention sequence $\{\mathcal{A}_n\}_{n\in 0\ldots N-1}$.
We know from Ref.~\onlinecite{Pollock2018} that the process tensor is the unique multi-linear map with this property.
We can thus conclude that the tensor network within the red shaded area in Fig.~\ref{fig:pt-tempo-network-a} (excluding the interventions $\mathcal{A}_0, \mathcal{A}_1, \ldots$) is a representation of the desired process tensor $\mathcal{T}_{0:N}$.

As we will see in section~\ref{sec:use-cases}, it is often advantageous to consider the slightly smaller part of this tensor network shown as the dotted area in Figs.~\ref{fig:pt-tempo-network}, which excludes the initial state $\rho^\mathrm{S}_0$ and the system propagators $\mathcal{V}_n$ and $\mathcal{V}^\prime_n$.
This corresponds to the process tensor $\tilde{\mathcal{T}}_{0:N}$ for an initially uncorrelated state and a total Hamiltonian $\tilde{H} = \hat{H}_\mathrm{I} + \hat{H}_\mathrm{E}$, without any contribution from the system Hamiltonian.
From this perspective the final state is then
\begin{align}
	\rho^\mathrm{S}(t_N)
	&= \mathcal{T}_{0:N}\left[ \mathcal{A}_{N-1}, \ldots, \mathcal{A}_1, \mathcal{A}_0 \right] \\
	&= P^\prime_{N-1} \tilde{\mathcal{T}}_{0:N}\left[ \tilde{\mathcal{A}}_{N-1}, \ldots, \tilde{\mathcal{A}}_1, \tilde{\mathcal{A}}_0 \right]\mathrm{,}
\end{align}
with $\tilde{\mathcal{A}}_n = \mathcal{V}_{n} \circ \mathcal{A}_n \circ \mathcal{V}^\prime_{n-1}$ and $\tilde{\mathcal{A}}_0: \rho^\mathrm{S} \mapsto \mathcal{A}_0 \left[ \mathcal{V}_{0} \left( \rho^\mathrm{S}_0 \right) \right]$.
This means that the evolution due to the system part of the Hamiltonian can be realized as a discrete set of interventions when the time steps $\delta t$ are chosen small enough such that the higher order terms of a second-order Trotterization can be neglected.

To compute $\tilde{\mathcal{T}}_{0:N}$ from the tensor network, Jørgensen and Pollock suggest to contract the network column by column with appropriate SVD sweeps during the process~\cite{Jorgensen2019}.
The computation time of this algorithm scales as $O(N \, \chi^3 d^6\, K_\mathrm{max})$, where $\chi$ is the bond dimension of the PT-MPO, $d$ is the Hilbert space dimension of the system, and $K_\mathrm{max} = \tau_\mathrm{cut} /\delta t$.

One improvement that can be made to this tensor network is implementing a so-called \emph{degeneracy checking} on the internal legs of $b_k$ tensors~\cite{Cygorek2017, Strathearn2018, minoguchi2019}.
This is where identical rows of the $b_k$ tensors are grouped together, reducing the bond dimensions of the internal legs of the tensor network.
The east--west legs can be grouped together by equal energy differences in the coupling operator eigenvalues, as only the energy difference affects the result of the influence functional.
The north--south legs require additionally degeneracy in the sum of eigenvalues as well as the difference.
For evenly spaced eigenvalues, like collective spin operators, this gives a best case reduction of the scaling of the computation time from $O(d^6)$ to $O(d^3)$.
In general, the reduction in numerical effort is related to the number of distinct eigenvalue differences in the system-bath coupling operator~\cite{Cygorek2017}.

\subsubsection*{Construction of PT-MPOs in \oqupy{}}
The PT-MPO construction with PT-TEMPO is handled by the \code{pt\_tempo\_compute()} function in \oqupy{}.
This takes, as its leading argument, a \code{Bath} object containing both the spectral density \(J(\omega)\)  of the environment and the operator \(\hat{S}\) coupling it to the system.
For each time step, influence tensors for the \code{Bath} are calculated and subsequently contracted to produce the PT-MPO.
The \code{ProcessTensor} object returned by \code{pt\_tempo\_compute()} contains the PT-MPO and can be held in memory or written to disk for later use.
Note that we use a \code{monospace font} throughout this article for expressions (classes, functions, and variable names) that are part of the \oqupy{} software package.

We now explain the computational parameters relevant to the PT-TEMPO method in \oqupy{}.
PT-TEMPO is a numerically exact method in the sense that no approximations are required in its derivation.
Error then only arises in its numerical implementation, controlled by a set of computational parameters.
The calculated dynamics can, in principle (i.e., with unlimited computational resource), be made as accurate as desired by tuning those parameters.

There are three main computational parameters to be decided.
The first is a memory cut-off $\code{tcut}\hat{=}\tau_\mathrm{cut}$ for the PT-MPO, set physically by the period of time $\tau_\mathrm{cut}$ over which non-Markovian effects persist via the environment.
This may be estimated by inspecting the decay of the environment auto-correlation function.
In practice, a suitable value is verified by repeating the calculation at longer \code{tcut} and checking the result is unchanged.

Second, a time step length $\code{dt}\hat{=}\delta t$ must be specified.
This must be short enough so as to avoid Trotter error in the discretization of the path integral for the system dynamics that is the TEMPO tensor network.
At the same time, \code{dt} fixes the grid upon which the calculation is performed, so also should be short enough to provide satisfactory resolution of the dynamics.

A precision \code{epsrel} completes the set of core computational parameters.
This sets the relative cut-off for singular value truncation when constructing the PT-MPO.
It must therefore be sufficiently small so as to avoid error resulting from the discard of physical correlations.
Note that PT-TEMPO runs at a fixed precision $\epsilon_{\text{rel.}}$ such that in the construction of the PT-MPO singular values smaller than $\epsilon_{\text{rel.}}$ relative to the largest singular value are discarded.
This is in contrast to other MPO methods where instead the bond dimension of the tensor is fixed and the precision varies.

Similar to the \code{tcut} parameter, to verify the suitability of chosen values of \code{dt} and \code{epsrel}, one may repeat the computation with more stringent (smaller) values and check the convergence of results.
Of course, smaller \code{dt} and \code{epsrel} (or longer \code{tcut}) typically imply a more resource intense calculation; in general a balance must be sought between the accuracy of the dynamics and computational cost.
The utility of the PT-TEMPO method is in the realization that results of sufficient accuracy can be obtained at a manageable computational cost for a wide range of problems. 
In \oqupy{}, the parameters \code{tcut}, \code{dt} and \code{epsrel} are encapsulated in a \code{TempoParameters} object and passed as an argument to \code{pt\_tempo\_compute()} at the stage of creating the PT-MPO.

As an example, we consider a spin-boson model with a coupling operator $\hat{S}_z$ (spin operator in Z-direction) and a spectral density of the form in Eq.\eqref{eq:spectral-density-power-law} with $\alpha=0.1$, $\zeta=1.0$, $\omega_c=1.0\,\ips$, at temperature \mbox{$T=1.0\,\ips \,(=7.64\,\mathrm{K})$}. 
Here and in the following, we have set $\hbar=k_\mathrm{B}=1$ and give all frequencies in units of \ips.
This choice of units is motivated by the fact that most physical examples discussed in Section~\ref{sec:use-cases} take place at this time scale.
We note, however, that there is no handling of units in \oqupy{}. Hence, all input and outputs of frequency and time are without unit and need to be interpreted by the user with respect to some (arbitrary) characteristic frequency.

In Fig.~\ref{fig:pt-tempo-results-a} we plot the computation time taken by \code{pt\_tempo\_compute} function with the parameters $\code{tcut}=2.0$, $\code{dt}=0.0625\,\ps$, and $\code{epsrel}=6.1\times10^{-5}$ for 32 time steps.
We see that the computation time to generate the process tensor scales with the system size and that it can be improved by the use of the previously discussed degeneracy checking.

\section{PT-MPO Use Cases}
\label{sec:use-cases}

In this section we discuss a number of use cases that are accommodated by the current version (v0.5) of the \oqupy{} package~\cite{OQuPy2022}.
In addition to the calculation of dynamics for a quantum system coupled to one or more structured environments~\cite{Strathearn2018, Gribben2021}, these include calculations of multi-time correlation functions~\cite{de_wit_in_preparation}, optimization of control protocols~\cite{Fux2021, butler_optimizing_2024}, and the dynamics of \textit{many-body} open quantum systems described by chains~\cite{fux_tensor_2023} or many-to-one networks of open quantum systems~\cite{fowler2022}.
The range of these examples demonstrate the versatility of the process tensor approach for extracting static and dynamic quantities from models of open quantum systems.

All of the use cases presented in this section apply pre-computed PT-MPO representations of environments for specific tasks.
We will always obtain such a PT-MPO by employing the PT-TEMPO method for Gaussian bosonic environments described in the previous Section~\ref{sub:pt-tempo}.
We note again, however, that the methods presented in this section are agnostic to the origin of a PT-MPO, and as such can also be readily applied to PT-MPOs for other types of environments that have been computed with other methods and software. 

The applicability of the methods presented in this section depends to a large extent on the complexity of the PT-MPOs that enter into the computation.
This complexity is typically reflected in the bond dimension of PT-MPOs.
The main restriction of the methods presented in this section is thus that for the environment of interest a sufficiently accurate PT-MPO with a manageable bond dimension ($\chi \lesssim 1000$) exists, and that it can be obtained with one of the known methods.
Although this is certainly not always the case, the literature on known methods to obtain PT-MPOs~\citePTcreation{} demonstrates that this is feasible for a large range of different highly relevant non-Markovian environments.

For every use case we outline the inner workings of the numerical method and show the results of an example, forgoing further discussion of the resulting physics.
For more details on the methods and physical discussion of results, we refer the reader to the research articles of the individual methods~\citeallmethods{}.
We note that all computation times quoted in this paper have been determined by running \oqupy{} on a \emph{single} core of an Intel i7 (8th Gen) processor. 

\subsection{Dynamics of non-Markovian open quantum systems}
\label{sub:simple-dynamics}

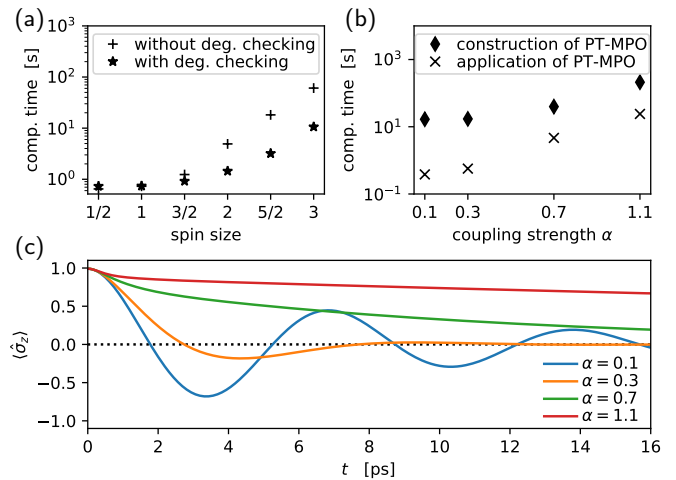
\begin{figure}
	\phantomsubfloat{\label{fig:pt-tempo-results-a}}%
	\phantomsubfloat{\label{fig:pt-tempo-results-b}}%
	\phantomsubfloat{\label{fig:pt-tempo-results-c}}%
	\begin{tikzpicture}[every node/.style={font=\sffamily}]
		\node (a) at (0,0) {\includegraphics[width=0.49\textwidth]{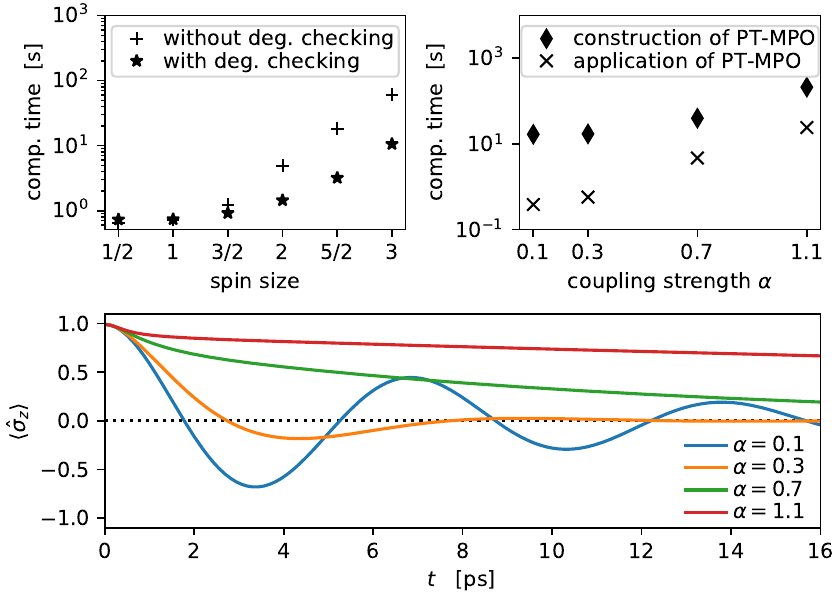}};
		\node [shift={(0.44,-0.20)}] at (a.north west) {(a)};
		\node [shift={(4.84,-0.20)}] at (a.north west) {(b)};
		\node [shift={(0.44,-3.23)}] at (a.north west) {(c)};	
	\end{tikzpicture}	
    \caption{\label{fig:pt-tempo-results}%
        (a)~Computation times for the construction of a PT-MPO using PT-TEMPO with/without degeneracy checking for different system sizes, with parameters described in the main text of Section~\ref{sub:pt-tempo}.
        (b)~Computation times for the construction and application of a PT-MPO for different coupling strengths of a biased spin-boson model described in the main text of Section~\ref{sub:simple-dynamics}.
        (c)~Dynamics of the biased spin-boson model for different coupling strengths.
    }
\end{figure}

Given a PT-MPO, constructed using the \code{pt\_tempo\_compute()} function or otherwise, single-time dynamics are computed in \oqupy{} following the introduction of a \code{System} object.
This describes the unitary evolution of the system (i.e., the system Hamiltonian) as well as coupling to any additional Markovian baths (Lindblad operators and rates).
The PT-MPO and \code{System} object are passed together with an argument prescribing the initial state of the system $\rho^\mathrm{S}_0$ to the \code{compute\_dynamics()} function, which performs the tensor network contractions required to yield the dynamics.
A \code{Control} object describing a set of control operations or interventions in the system dynamics may also be provided at this stage.

The result returned by \code{compute\_dynamics()} is a \code{Dynamics} object.
This contains a list of system states for each time step from which, for example, the time dependent expectation values of system observables may readily be derived.
Note that the same approach can be used in the case of coupling to multiple non-Markovian environments described by separate PT-MPOs; see Section~\ref{sub:multi-environments}.

To provide an example, we consider the quench dynamics of a simple spin-boson model with a Ohmic spectral density at different coupling strengths.
For this, the overall Hamiltonian is of the form of Eq.~\eqref{eq:tempo-hamiltonian}, with $H^\mathrm{S} = \frac{\Omega}{2} \hat{\sigma}_x$, coupling operator $\hat{S} = \frac{1}{2}\hat{\sigma}_z$.
We assume the spectral density $J(\omega)$ in the form of Eq.~\eqref{eq:spectral-density-power-law}, where $\Omega=1.0\,\ips$, $\zeta=1.0$, $\omega_c=5.0\,\ips$, and temperature $T=0$.
Figure~\ref{fig:pt-tempo-results-c} shows the dynamics of $\langle \sigma_z(t) \rangle$, starting in the spin-up state $\ket{\uparrow}$, for different coupling strengths $\alpha$.
For each coupling strength $\alpha$ we first use the \code{pt\_tempo\_compute()} function to generate a PT-MPO, and then apply the \code{compute\_dynamics()} function to obtain the dynamics.
In Fig.~\ref{fig:pt-tempo-results-b} we show the computation times for those calculations using $\code{tcut}=4.0\,\ps$, $\code{dt}=0.0625\,\ps$, $\code{epsrel}=6.1\times10^{-5}$ for $N=256$ time steps.

In Fig.~\ref{fig:pt-tempo-results-b} we can observe clearly that \emph{applying} a PT-MPO takes much less time than \emph{constructing} a PT-MPO.
This is because during the creation of the PT-MPO, the PT-TEMPO method identifies and removes negligible correlations, which constitutes the bulk of the overall computation.
This is particularly useful when one is interested in a large set of different multi-time correlations or the dynamics for many different system Hamiltonians, because in such cases the difficult creation of the PT-MPO has to performed only once, while the repeated application of it can be performed much more quickly.
This fact will be used extensively in the following two use cases on multi-time correlations and the optimization of control protocols.

\subsection{Multi-time correlations of non-Markovian open quantum systems}
\label{sub:multi-time-correlations}
The calculation of multi-time correlations for open quantum systems often involves application of the quantum regression theorem, which relies on making a Markovian approximation~\cite{Breuer2002}.
However, there exist many scenarios---from exciton dynamics in quantum dots to natural light-harvesting---where such an approximation is not sufficient~\cite{Chirolli2008, DeVega2017, wang2019, cao2020}.
Moreover, as mentioned in the introduction, even if a Markovian description correctly captures the reduced dynamics of the system, an accurate description of multi-time observables is not guaranteed~\cite{guarnieri2014, milz_quantum_2021}.
The PT-MPO, on the other hand, not only encodes the evolution of the density matrix but also all multi-time correlations.
These are accessed by simply inserting the operators at the corresponding times as control operations, depicted in Fig~\ref{fig:pt-tempo-network-d}.

Figure~\ref{fig:pt-tempo-network-d} shows the three-time correlation
\begin{equation}
R = \Tr \left[ \hat{C}(\tau_3) \hat{A}(\tau_1) \rho_0 \hat{B}(\tau_2) \right]
\end{equation}
with $\tau_1=t_0$, $\tau_2=t_2$, and $\tau_3=t_4$, where $\hat{X}(\tau)=U^\dagger(\tau) \hat{X} \hat{U}(\tau)$ denotes system operators in the Heisenberg picture with respect to the \emph{total} Hamiltonian.
Also, $\rho_0$ is the total initial state and $\Tr$ denotes the trace over the total (i.e. system \emph{and} environment) Hilbert space.
The control operations that need to be inserted into the process tensor at the corresponding times are the left and right acting superoperators~\footnote{Although these inserted ``control operations'' are not CP-maps, this procedure yields nonetheless correct and numerically exact results.}
$\mathcal{A}^\mathrm{L}[\rho] = \hat{A} \rho$, $\mathcal{B}^\mathrm{R}[\rho] =\rho \hat{B}$, and $\mathcal{C}^\mathrm{L}[\rho] = \hat{C} \rho$.
After the last control operation we trace over the system Hilbert space.
Inserting a left or right acting system operator and tracing over the system is equivalent to computing the expectation value of that operator.
Therefore, multi-time correlations for various final times can be computed in a single run of the simulation.
To vary any of the earlier time arguments, e.g. for $i$ time steps, the simulation is repeated $i$ times, shifting the position of the earlier control operations accordingly.
As explained in Section~\ref{sub:simple-dynamics}, since the process tensor is independent of the system propagators and control sequence, the same PT-MPO can be applied repeatedly, which drastically decreases the necessary computation time.

We note that the set of multi-time expectations that can be calculated this way require that the set of operators acting to the left and those acting to the right are separately time ordered.
There is no restriction on how the times of left-acting and right-acting operators are related, however one cannot calculate arbitrary out-of-time-order-correlations~\cite{Swingle2018}.

In \oqupy{}, multi-time correlations are computed with the function \code{compute\_correlations\_nt()}.
This requires a \code{System} and \code{ProcessTensor} object, together with three lists specifying the operators, the times at which they should be applied and whether each operator should be applied to the left or right of the (total) density matrix.

\begin{figure}
	\phantomsubfloat{\label{fig:n-time-corr-a}}%
	\phantomsubfloat{\label{fig:n-time-corr-b}}%
	\phantomsubfloat{\label{fig:n-time-corr-c}}%
	\phantomsubfloat{\label{fig:n-time-corr-d}}%
    \includegraphics[width=0.45\textwidth]{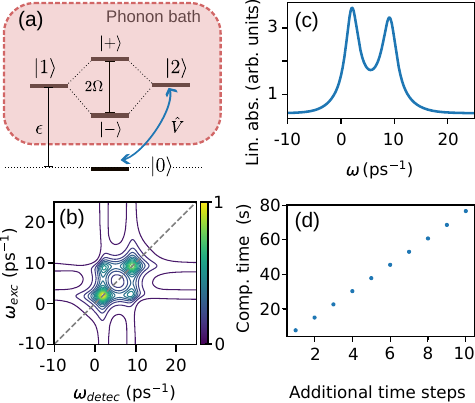}
    \caption{\label{fig:n-time-corr}%
        (a)~Sketch of model employed in the multi-time correlations use case in Eq.~\eqref{eq:3ls}; $\epsilon=5.0\,\ips$, $\Omega=2.0\,\ips$.
        (b) Corresponding 2D spectrum calculated from the four-time correlation functions in Eq.~\eqref{eq:R} at \mbox{$T=13.0\,\ips\,(=100\,\mathrm{K})$}.
        (c)~Corresponding linear absorption spectrum at \mbox{$T=13.0\,\ips$}.
        (d)~Scaling of the computation time for calculating the four-time correlation function $R_4(\tau_1, \tau_4)$ when adding additional time steps to $\tau_1$.
        Note that the same PT-MPO was used for each data point, which took 44~s to compute with $\code{tcut}=50.0\,\ps$, $\code{dt}=0.1\,\ps$ and $\code{epsrel}=10^{-6}$.
    }
\end{figure}

To demonstrate the utility of the code, we apply it to a three-level system and simulate a linear absorption and 2D electronic spectroscopy (2DES) measurement~\cite{de_wit_in_preparation}.
2DES is a non-linear spectroscopy technique that uses a series of laser pulses to probe energy and charge transport on a femtosecond timescale.
The resulting signal is commonly visualized as a 2D spectrum that correlates the excitation and detection frequencies excited by the lasers, see Fig.~\ref{fig:n-time-corr-b}.
Because of its sensitivity to complex phases, 2DES is particularly useful for investigating the coherent dynamics of energy excitations in real time.
For a more detailed overview, we refer the reader to Refs.~\onlinecite{cho2008, collini2021}.

As shown in Fig.~\ref{fig:n-time-corr-a}, our model is a three-level system with a linear coupling to a phonon bath.
The total Hamiltonian has the form of Eq.~\eqref{eq:tempo-hamiltonian}, where the system Hamiltonian and coupling operator are
\begin{equation}
    \label{eq:3ls} 
    \hat{H}_S = (\epsilon + \lambda)\pqty{ \ketbra{1}{1} + \ketbra{2}{2}} + \Omega\pqty{\ketbra{1}{2} + \mathrm{H.c.}},
\end{equation}
and $\hat{S} = \pqty{\ketbra{1}{1} - \ketbra{2}{2}}$.
This Hamiltonian describes an excitonic dimer embedded in a vibrational environment with ground state $\ket{0}$ and monomer states $\ket{1}, \ket{2}$.
Each monomer state has a bare energy $\epsilon + \lambda$, where $\lambda$ represents the bath reorganization energy (defined below).
The coupling $\Omega$ creates two delocalized exciton states $\ket{\pm} = \frac{1}{\sqrt{2}} \pqty{\ket{1} \pm \ket{2} }$ with energies $E_{\pm} = \epsilon + \lambda \pm \Omega$.
Since the dynamics of the type of molecules studied with 2DES generally take place on a picosecond timescale~\cite{wang2019}, we set $\epsilon=5.0\,\ips$ and $\Omega = 2.0\,\ips$.
The spectral density $J(\omega)$ is given by Eq.~\eqref{eq:spectral-density-power-law} with $\zeta=1.0$, $\alpha=0.1$ and $\omega_c=3.04\,\ips$, and leads to the reorganization energy $\lambda = \int_0 ^\infty \frac{1}{\omega} J(\omega)\, \dd\omega = 2\,\alpha\,\omega_c.$ 

A 2DES signal can be modeled by a series of four-time correlation functions that represent all possible light-matter interaction pathways.
Here we will consider the following four correlation functions:
\begin{flalign}
    R_1 &= \Tr \big[ \hat{V}(\tau_4)\hat{V}(\tau_1)\rho_0\hat{V}(\tau_2)\hat{V}(\tau_3) \big] \nonumber \\
    R_2 &= \Tr \big[\hat{V}(\tau_4)\hat{V}(\tau_2)\rho_0 \hat{V}(\tau_1)\hat{V}(\tau_3) \big] \nonumber \\
    R_3 &= \Tr \big[\hat{V}(\tau_4)\hat{V}(\tau_3)\rho_0 \hat{V}(\tau_1)\hat{V}(\tau_2) \big] \nonumber \\
    R_4 &= \Tr \big[\hat{V}(\tau_4)\hat{V}(\tau_3) \hat{V}(\tau_2)\hat{V}(\tau_1) \rho_0 \big],  
    \label{eq:R}
\end{flalign}
where $\hat{V} = \ketbra{0}{2} + \ketbra{2}{0}$. 
To plot a 2D spectrum, each correlation function is Fourier transformed with respect to the first ($\tau_2 - \tau_1$) and last ($\tau_4 - \tau_3$) time delay, corresponding to an excitation ($\omega_\mathrm{exc}$) and detection ($\omega_\mathrm{detec}$) frequency axis.
We then sum together the real parts of each pathway to obtain the total spectrum, which is shown in Fig.~\ref{fig:n-time-corr-b}.
Here for simplicity, we have set the second time delay $\tau_3 - \tau_2$ to zero.

Simulating a linear absorption spectrum for our model requires computing the two-time correlation function $\Tr \big[ \hat{V}(\tau_2)\hat{V}(\tau_1)\rho_0\big] $.
As with the two-dimensional case, the spectrum is given by the Fourier transform of this function with respect to $\tau_2 - \tau_1$, and is shown in Fig.~\ref{fig:n-time-corr-c}.
The spectrum contains two peaks corresponding to the $\ket{+}$ and $\ket{-}$ states excited by the dipole operator $\hat{V}$.
The 2D spectrum (Fig.~\ref{fig:n-time-corr-b}) contains two diagonal peaks, reflecting the same transition frequencies observed in the linear absorption spectrum.
Furthermore, two off-diagonal peaks that cross-correlate the two frequencies can be observed; these signify the presence of the electronic coupling $\Omega$ between the excited states.

To illustrate how this computation scales with the range of time steps, we first consider the correlation function $R_4(\tau_4)$ in Eq.~\eqref{eq:R}.
As a function of the final time argument $\tau_4$ only, the simulation runs once and took 8~s to compute.
If we additionally vary any of the earlier time arguments, e.g. $\tau_1$ in $R_4(\tau_1,\tau_4)$ over $i$ time steps, the simulation is repeated $i$ times.
As shown in Fig.~\ref{fig:n-time-corr-d}, the computation time therefore scales linearly with the number of additional time steps.
The 2D spectrum in Fig.~\ref{fig:n-time-corr-b} was computed for 40 time steps each in $\tau_1$ and $\tau_4$ and consists of the four correlation functions in Eq.~\eqref{eq:R}, giving a total computation time of 21~minutes.

\subsection{Optimization of control protocols}
\label{sub:optimization}

As mentioned in the end of Section~\ref{sub:simple-dynamics}, the fact that the PT-MPO does not depend on the system Hamiltonian allows one to efficiently determine optimal control protocols for non-Markovian open quantum systems~\cite{Fux2021, butler_optimizing_2024}.
Optimal control involves defining an objective function to quantify the success of a given protocol, and then maximizing this function over the set of controls.
A common scenario is for the protocols to correspond to different time-dependent system Hamiltonians, and for the value of the objective function to be determined by the final state.
The optimization can be done by first computing the PT-MPO of the given environment interaction, and then repeatedly applying different time-dependent system Hamiltonians. 
This has the advantage that each trial system Hamiltonian can be applied with minimal computational effort to the same pre-computed PT-MPO.

A further advantage of the PT-MPO approach is that it provides a natural and efficient way to compute the gradient of the objective function with respect to the parameterization of the system Hamiltonian.
This allows one to use the gradient in the optimization process, drastically reducing the computation time required. 
In this section, we outline how OQuPy is used to calculate the gradient of an objective function with respect to a set of parameters. 

The basis for the gradient calculation is the observation that the process tensor is a multi-linear map from the set of system propagators---or, more generally, system operations---to the final state.
This implies that the derivative of the final state with respect to a particular system propagator is the diagram for the final state, such as Fig. \ref{fig:pt-tempo-network-c}, with that propagator omitted.
It can be constructed by combining two partial diagrams, one obtained by contracting the network starting from the bottom, and one starting from the top, in each case stopping just before the omitted propagator.
Furthermore, the partial diagrams required to compute the derivatives with respect to all the propagators, i.e. the gradient, are computed in the course of one forwards-in-time propagation, from bottom to top, and one backwards-in-time propagation, from top to bottom.
The forward propagation would, in any case, be needed to compute the objective function, so the additional cost of this process is that of the backwards propagation and contractions of the partial diagrams. 

In OQuPy this calculation is focused on the problem of computing the gradient of an objective function, defined in terms of the final state, $Z(\rho_f)$, with respect to some parameterization of the time-dependent system Hamiltonian. We define a set of parameters $\{c_\alpha\}$, with $\alpha=0,\ldots,M$ which appear in the system Hamiltonian, and take the values $c_\alpha (t_n)=c_\alpha^n$ at the $n^{th}$ time step. The derivative of the objective function with respect to $c_\alpha^n$ is given by the chain rule,
\begin{equation}
\frac{\partial Z}{\partial c_\alpha^n}= \underbrace{\sum_{i,j,k}^{d_{H_S}^2} \:
\underbrace{\frac{\partial Z}{\partial \rho_f^i} \quad
\frac{\partial\rho_f^i}{\partial \mathcal{V}^{jk}_n}}_{\substack{\text{\code{compute\_gradient}} \\ \text{\code{\_and\_dynamics()}}}}
\underbrace{\frac{\partial \mathcal{V}^{jk}_n}{\partial c_\alpha^n \vphantom{\rho^i_f} }}_{\substack{\text{\code{Parameterized}} \\
\text{\code{System}}}}}_{\text{\code{state\_gradient()}}},\label{eq:dzdc}
\end{equation}
where $\rho_f$ is the final state and $\mathcal{V}_n$ the propagator(s) for the time step $n$. The second factor in Eq. (\ref{eq:dzdc}) can be computed for all $n$, as discussed above, using one forwards and one backwards propagation, and combined with the remaining terms to compute the gradient, $\frac{\partial Z}{\partial c_\alpha^n}$ for all $n,\alpha$.

There are three steps in computing the gradient of an objective function of a final state with respect to a set of parameters in OQuPy. 
The first step is to define a system via the \code{ParameterizedSystem} object.
This object inherits from \code{BaseSystem} and represents a Hamiltonian which depends on $M$ parameters $\{c_0,c_1,\ldots \}$.
Instantiation requires the user to supply a function which takes values for these parameters and returns the Hamiltonian operator.
The \code{ParameterizedSystem} class then handles the calculation of the propagators $\mathcal{V}_n$ and propagator derivatives $\frac{\partial \mathcal{V}_n}{\partial c^n_\alpha}$ during the gradient calculation via the \code{get\_propagators()} and \code{get\_propagator\_derivatives()} methods.
Alternatively, the user can provide a function which returns the propagator derivatives. 

The simplest calculation one can do with a \code{ParameterizedSystem} object is to compute the dynamics.
This can be done by calling the routine \code{compute\_gradient\_and\_dynamics} and providing values of the parameters, $c_\alpha^n$ for all time steps.
An additional detail is that the calculations use a second-order Trotter splitting, so that, as shown in Fig.~\ref{fig:pt-tempo-network}, propagation over a full time step involves two propagators, each over half a time step, with one, the pre-propagator $\mathcal{V}_n$ applied to the corresponding input leg of the PT-MPO, and one, the post-propagator $\mathcal{V}^\prime_n$, to the output leg.
The computation of the dynamics thus requires the values of the parameters on both halves of every time step.
The results are accurate up to second order in \code{dt} provided the parameters are continuous within each full time step, i.e. the difference in the Hamiltonian between the first and second halves of each step is $O(\code{dt})$.
Note there is no requirement for continuity between full time steps, so that discontinuous Hamiltonians can be treated correctly.

To compute the gradient of an objective function, the \code{ParameterizedSystem} object, along with the PT-MPO, initial state, and values for the control parameters, can be passed to the function \code{state\_gradient()}.
The objective function can be specified in two ways. For the case of a linear objective function, $Z=\sum_{ij} (A)_{ij}(\rho_f)_{ij}$, the first factor in Eq. (\ref{eq:dzdc}) is $\frac{\partial Z}{\partial \rho_f} = A$, which can be provided to \code{state\_gradient()} as an array.
An example arises for state transfer where we seek to reach a pure target state, $\rho_t=\ketbra{\sigma}{\sigma}$, at the final time.
The fidelity $\mathcal{F} = \bra{\sigma} \rho_f \ket{\sigma}$ is then a linear objective function with $A=\rho_t^T$.
For nonlinear objective functions, the user passes a function to \code{state\_gradient()}.
This will be called with the computed final state as an argument and should return $\frac{\partial Z}{\partial \rho_f}$ evaluated for that state. 

\begin{figure}
	\phantomsubfloat{\label{fig:gradient-a}}%
	\phantomsubfloat{\label{fig:gradient-b}}%
	\includegraphics{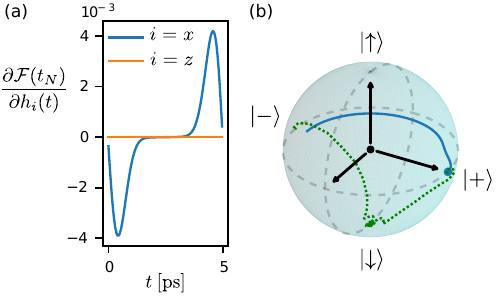}
	\caption{\label{fig:gradient}%
		(a)~Derivative of the fidelity at the final time $t_N$ with respect to the parameters at each time-step $h_x(t)$ and $h_z(t)$ for initial controls.
		(b)~Dynamics of states on Bloch sphere under initial (blue, solid) and optimized (green, dotted) controls.}
\end{figure}

The \code{state\_gradient()} function handles the forwards and backwards propagations as well as the application of the chain rule. It first passes the arguments to the \code{compute\_gradient\_and\_dynamics()} function, yielding the tensors $\frac{\partial Z}{\partial \rho_f^i} \frac{\partial\rho_f^i}{\partial \mathcal{V}^{jk}_n}$.
These tensors are then passed to the \code{chain\_rule()} function, where they are combined with the propagators and propagator derivatives for each half time step.
The result is a list of derivatives of the objective function with respect to the control parameters at each half time step.
The final dictionary returned by \code{state\_gradient()} function includes a \code{Dynamics} object, the gradient tensors  $\frac{\partial Z}{\partial \rho_f} \frac{\partial\rho_f}{\partial \mathcal{V}^{jk}_n}$, and the derivatives at each half time step $\frac{\partial{Z}}{\partial{c^n_\alpha}}$. 

Fig. \ref{fig:gradient-a} shows an example of the gradient computation for a two-level system coupled to a Gaussian bosonic bath, with a super-Ohmic spectral density $J(\omega)$ given by Eq.~\eqref{eq:spectral-density-power-law} with $\zeta=3.0$, $\alpha=0.126$, and $\omega_c = 3.04\,\ips$.
We consider an initial state $\rho_0=\ketbra{+}{+}$, and take the objective function to be the fidelity $\mathcal{F}(\rho_f, \rho_t)$ to a target state $\rho_t=\ketbra{-}{-}$.
The system Hamiltonian and coupling operator are $\hat{H}_\mathrm{S}=h_x (t) \hat{\sigma}_x + h_z (t) \hat{\sigma}_z$ and $\hat{S}=\frac{\hat{\sigma}_z}{2}$, and the dynamics is simulated with $N=100$ time steps.
The gradient is shown for the case of  control fields $h_x(t)=0$ and $h_z(t)=\pi / t_N$, which are the optimal controls in the absence of the environment.
The dynamics of the state on the Bloch sphere under this set of controls is depicted by the blue solid line in Fig.~\ref{fig:gradient-a}.

The \code{state\_gradient()} function can be used within a numerical optimization routine to determine optimal control protocols.
The green dotted curve in \ref{fig:gradient-b} shows the trajectory for optimal controls determined in this way, using the L-BFGS algorithm implemented in the \code{scipy.optimize.minimize()} function~\cite{virtanen_scipy_2020}.
This optimization was done using bounds $|h_x| \leq h^\text{max}_x=5.0\pi\,\ips$ and $|h_z| \leq h^\text{max}_z=1.0\pi\,\ips$, which restrict the speed of the unitary evolution.

For this optimal control, the fidelity of the final state with respect to the target state is $0.9991$.
From the dynamics of the Bloch vector it can be seen that the increase in fidelity over the unoptimized solution arises because the state is transferred to an intermediate state $\ket{\downarrow}$, which is not subject to decoherence~\cite{basilewitsch_optimally_2020}.
A different mechanism can appear with smaller bounds on the control fields, which prohibit access to that decoherence-free subspace.
In that case, it has been shown that for larger process durations there can be an increase in the fidelity \cite{butler_optimizing_2024} due to information previously lost to the environment being restored by the optimization.
The optimization utilizes information back-flow from the environment by maximizing the non-Markovianity of the map.

\subsection{Open system dynamics with multiple environments}
\label{sub:multi-environments}
It can sometimes be sensible, and even necessary, to express the external influences on a system in terms of multiple environments.
A typical case in physics is that of an optically active system that is also strongly coupled to its vibrational environment.
For example, in systems studied for light harvesting and energy transfer the molecular vibrational degrees of freedom can play a key role~\cite{hedley2017light,adronov2000light, felip2016chameleonic, burzuri2016sequential,scholes2011lessons}.
Strong coupling to vibrations has also been shown to have a distinct effect on the optical properties of even relatively simple systems~\cite{Gribben2022,Maguire2019}; the combined effect of multiple environments in this way is termed \emph{non-additive}.
Another example of non-additivity occurs when considering the combined effect of the leads and vibrational environments of molecular nanojunctions~\cite{thomas2019understanding, sowa2020beyond}; here using an additive treatment can even lead to a violation of the Carnot bound on efficiency~\cite{mcconnell2022strong}. 

\begin{figure}
   	\phantomsubfloat{\label{fig:mult-env-construction-a}}%
    \phantomsubfloat{\label{fig:mult-env-construction-b}}%
    \includegraphics[width=\linewidth]{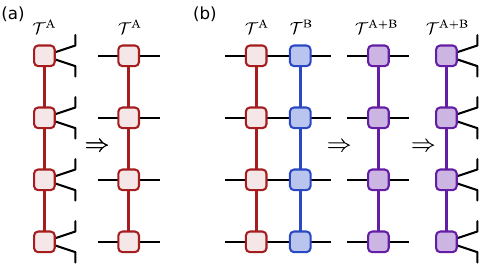}
    \caption{\label{fig:mult-env-construction}%
        Tensor network sketch of the addition of two PT-MPOs.
        (a)~Suitable reshaping of an PT-MPO.
        (b)~Contraction of two PT-MPOs to a single PT-MPO that represents the corresponding joint environment.
    }
\end{figure}

Typically PT-MPOs containing information on multiple environments can be much more costly to construct and store; this is especially true when the environments couple via non-commuting system operators.
Fortunately, we can avoid this cost by making use of the additive property of PT-MPOs, namely that two PT-MPOs can be contracted together to yield a single process tensor with the effect of both environments.
This allows us to construct one PT-MPO for each environment and store them separately, only combining them when necessary.
This step corresponds to a standard MPO--MPO multiplication as depicted in Figure~\ref{fig:mult-env-construction}.
However, combining the entirety of each environment's PT-MPO typically offers little to no improvement over constructing the single large object in the first place.
We can avoid this in certain calculations by only combining the tensors necessary at each step.
For example, when computing an observable at a single time step $t_n$ we can combine the PT-MPOs one time step at a time absorbing the tensor of each previous time step until we reach $t_n$.
This avoids the extra memory requirements in combining PT-MPOs and the computational overhead in compressing the resulting object.

\begin{figure}
	\centering
	\includegraphics[width=0.40\textwidth]{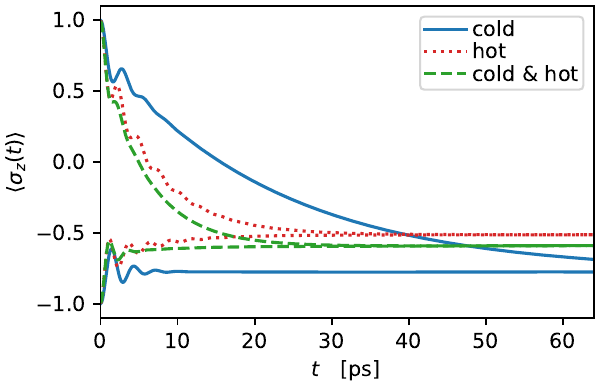}
	\caption{\label{fig:multi-env-results} 
        Evolution of a spin coupled to a cold bath, a hot bath, and to both baths simultaneously.
        We show the dynamics for two different initial states, $\ket{\uparrow}$ and $\ket{\downarrow}$.
    }
\end{figure}

As a demonstration we consider a two-level system coupled to both a hot and cold bath of bosons.
The total Hamiltonian is analogous to Eq.~\eqref{eq:tempo-hamiltonian}, but with two interaction and two environment parts $H^\mathrm{I}_a$, $H^\mathrm{E}_a$, and $H^\mathrm{I}_b$, $H^\mathrm{E}_b$, with bosonic environment annihilation operators $\hat{a}_k$ and $\hat{b}_k$, respectively.
We choose the system Hamiltonian $\hat{H}_\mathrm{S} = \frac{\epsilon}{2} \hat{\sigma}_z + \frac{\Omega}{2} \hat{\sigma}_x$, with $\epsilon=2.0\,\ips$, and $\Omega=1.0\,\ips$.
Both environments have the same Ohmic spectral density $J(\omega)$ given by Eq.~\eqref{eq:spectral-density-power-law}, with $\zeta=1.0$, $\alpha=0.16$, and $\omega_c=1.0\,\ips$.
The environments are distinguished by the spin component to which they couple and their temperature.
The cold bath, at temperature $T_\text{cold}=0.8\,\ips \,(=6.11\,\mathrm{K})$, couples via $\hat{S}_a=\hat{\sigma}_z/2$, while the hot bath, at temperature $T_\text{hot}=1.6\,\ips\,(=12.22\,\mathrm{K})$, couples via $\hat{S}_b=\hat{\sigma}_x/2$.
The dynamics of this model for initial states spin-up and spin-down are plotted in Figure~\ref{fig:multi-env-results} and the combined effect of the baths, when compared with their separate impact, can be seen clearly.

The combination of multiple PT-MPOs corresponds to standard tensor network operations but, as mentioned, this is often very demanding; in the above example the PT-MPOs had a maximum bond dimensions of $\chi_\text{cold}=35$ and $\chi_\text{hot}=41$.
\oqupy{} is instead able to make use of the more efficient calculation of dynamics introduced earlier in this section.
Doing so requires simply providing the \code{compute\_dynamics()} function with a list of PT-MPOs of the multiple environments and, provided they are compatible, the dynamics step-by-step will be computed.
The construction of each PT-MPO for Fig.~\ref{fig:multi-env-results} took $40\,\mathrm{s}$ (cold) and $51\,\mathrm{s}$ (hot) and the dynamics for combined simulation took $8\,\mathrm{s}$ compared with $3\,\mathrm{s}$ and $5\,\mathrm{s}$ for the separate cold and hot bath dynamics respectively.

In this section we highlighted the simplicity with which the dynamics of a system coupled to multiple environments can be computed with \oqupy{}.
The implementation allows for great flexibility in considering the effect of arbitrary combinations of PT-MPOs on the dynamics of the system.
The above example demonstrates this with two PT-MPOs for Gaussian bosonic baths, but---like all methods in this section---is also readily applicable to combinations of PT-MPOs that represent environments of different nature created with other methods and other software.

\subsection{Chains of open quantum systems}
\label{sub:pt-tebd}

\begin{figure}
	\phantomsubfloat{\label{fig:pt-tebd-approach-a}}%
	\phantomsubfloat{\label{fig:pt-tebd-approach-b}}%
	\includegraphics[width=0.45\textwidth]{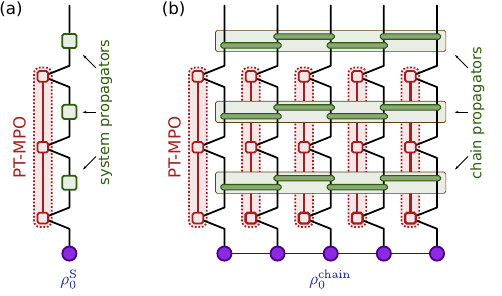}
	\caption{\label{fig:pt-tebd-approach}%
        Tensor network for the evolution of (a) a single system and (b) a chain of systems coupled to their local environment.
        Panel~(a) shows the state of a system after three time steps evolved by system propagators under the influence of an environment (represented by an PT-MPO), starting from the initial state $\rho^\mathrm{S}_0$.
        Panel~(b) shows the state of a chain of systems after three time steps under the influences of locally coupled environments, starting from an initial state represented as an MPS in Liouville space.
        The dark green rectangles are Trotterized chain propagators.
    }
\end{figure}

The methods and examples considered so far always involved a single (small) system coupled to one or more environments.
There is, however, a range of interesting physical scenarios where the system of interest is a many-body system that couples to one or more environments~\cite{Blanter2000, Agrait2003, Giazotto2006, Losego2012, Widawsky2013}.
Such scenarios are of importance for fundamental research, such as the study of strong coupling quantum thermodynamics ~\cite{Nicolin2011, Horodecki2013, Vinjanampathy2016, Seifert2016, Binder2019, Talkner2020}, as well as technological and biological applications~\cite{Bose2003, Wojcik2005, Engel2007, Lambert2013, Motlagh2014, Mitchison2019, cao2020}.
The method presented in this section enables the computation of the dynamics and multi-time correlations of chains of general open quantum systems~\cite{fux_tensor_2023}.
It is a combination of the time evolving block decimation (TEBD) method~\cite{Vidal2004} in Liouville space with PT-MPOs on each site to include the influence of locally coupled environments  (see Fig.~\ref{fig:pt-tebd-approach}).
We call this tensor network method \emph{PT-MPO augmented TEBD} (or short \emph{PT-TEBD}) because the TEBD method is augmented with one additional leg for each environment connecting the site with its PT-MPO.
This augmented leg encodes the correlations of the site with its environment.
Figure~\ref{fig:pt-tebd-approach-b} shows the tensor network for a first order Trotterized PT-MPO augmented TEBD tensor network.
This tensor network is suitable for simulating a chain of locally interacting systems where each site may couple strongly to its individual environment.

The PT-TEBD method has three convergence parameters.
These are (1) the Trotterization time step $\delta t$ (which should be the same as the time step chosen for the PT-MPO), (2) the Trotterization order $O_\mathrm{TEBD}$ (only 1$^\mathrm{st}$ and 2$^\mathrm{nd}$ are currently implemented), and (3) the relative cutoff $\epsilon_\mathrm{TEBD}$ for the SVD truncation along the spatial direction of the chain. 
The method is limited to chains whose state can be well approximated by an matrix product state (MPS) of some finite bond dimension $\xi$, as well as environments whose process tensor can be well approximated by an MPO of bond dimension $\chi$.
The computational complexity is then dominated by performing the singular SVD involved in compressing the spatial MPS after the application of the system propagators.
In the worst case the largest matrices involved are of the dimension $(\xi \chi d^2) \times (\xi \chi d^2)$, where $d$ is the Hilbert space dimension of a single site.
In many cases this can be reduced to $(\eta d^2) \times (\eta d^2)$ with $\xi \lesssim \eta \leq \xi\chi$ by a careful choice of contraction and SVD order (see appendix A of Ref.~\onlinecite{fux_tensor_2023}).
The overall simulation of an $N$ site chain for $K$ time steps thus takes $O(N K \eta^3 d^6)$ operations.
This algorithm is---like the canonical TEBD algorithm---well suited for parallel computing, since each pair of neighboring sites can be evolved separately.

\begin{figure}
	\includegraphics[width=0.45\textwidth]{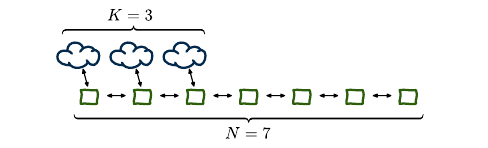}
	\caption{\label{fig:pt-tebd-example}%
		Sketch of an interacting chain of systems (green squares) with length $N=7$ coupling to local environments (blue clouds) on the first $K=3$ sites.  
	}
\end{figure}

To demonstrate the capabilities of this method we present the dynamics of an anisotropic XY-model where the first $K$ of $N$ sites each couple strongly to an environment (see sketch in Fig.~\ref{fig:pt-tebd-example}).
The chain Hamiltonian is of the form
\begin{equation}\label{eq:xyz-spin-chain}
	\hat{H}_\mathrm{XY} =  \sum_{n=1}^{N} \epsilon \hat{s}^z_n + \sum_{n=1}^{N-1} \left[
		(J - \eta) \hat{s}^x_n \hat{s}^x_{n+1} + (J + \eta) \hat{s}^y_n \hat{s}^y_{n+1}
		\right] \mathrm{,}
\end{equation}
with spin-1/2 operators $\hat{s}^\gamma_n = \hat{\sigma}^\gamma_n/2$ at site~$n$, onsite energy $\epsilon = 1.0\,\ips$, coupling strength $J=1.0\,\ips$ with anisotropy $\eta=0.04\,\ips$.
In \oqupy{} this chain Hamiltonian is encoded in a \code{SystemChain} object.
As an environment we choose a bosonic bath with an Ohmic spectral density $\zeta=1$, $\alpha=0.32$, $\omega_c=1.0\,\ips$ (see Eq.~\eqref{eq:spectral-density-power-law} and coupling operator $\hat{S} = \hat{\sigma}_z/2$. 
The corresponding PT-MPO has been computed using the PT-TEMPO algorithm with convergence parameters $\code{tcut}=4.0\,\ps$, $\code{dt}=2^{-4}\,\ps$, and $\code{epsrel}=2^{-16}$.
For the PT-TEBD algorithm we use $\epsilon_\mathrm{TEBD}=2^{-16}$ and $O_\mathrm{TEBD}=2$.
The parameters are collected in a \code{PtTebdParameters} object and passed into a \code{PtTebd} object together with the above \code{SystemChain} instance and a list of process tensors.  
We choose the state $\ket{\uparrow\downarrow\downarrow\cdots\downarrow}$ as the initial spin chain state, and are interested in the time evolution of $\bra{\uparrow} \rho^{(n)}(t) \ket{\uparrow}$ of each spin $n$. 

\begin{figure}
	\phantomsubfloat{\label{fig:pt-tebd-results-a}}
	\phantomsubfloat{\label{fig:pt-tebd-results-b}}
	\hspace*{-.25cm}%
	\begin{tikzpicture}[every node/.style={font=\sffamily}]
		\node (a) at (0,0) {\includegraphics[width=0.49\textwidth]{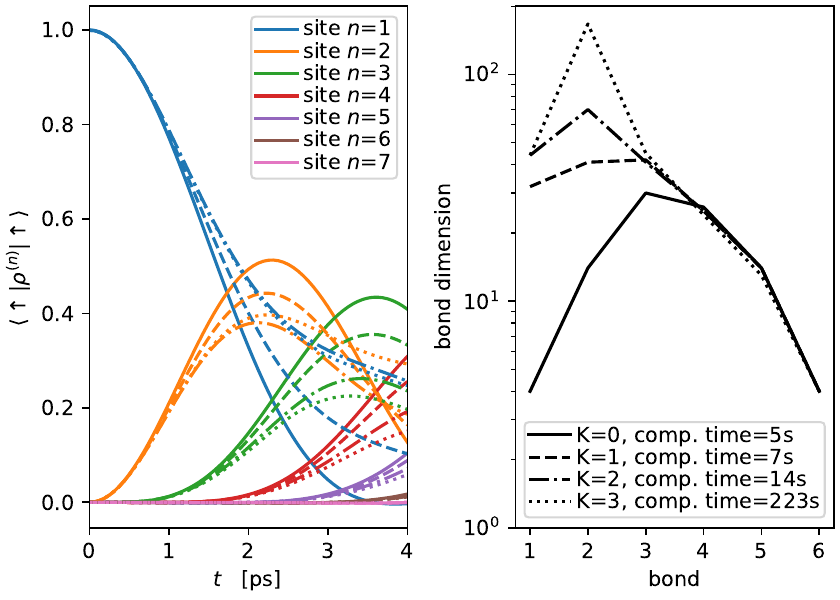}};
		\node [shift={(0.44,-.03)}] at (a.north west) {(a)};
		\node [shift={(4.84,-.03)}] at (a.north west) {(b)};
	\end{tikzpicture}		
	\caption{\label{fig:pt-tebd-results}%
        \oqupy{} results for the anisotropic XY-model with $N=7$ and $K=0\ldots3$.
        (a)~Evolution of the chain after the quench at $t=0$.
        (b)~Maximal bond dimension of the augmented MPS during the entire evolution at different bonds.
        The different line styles in panel (a) correspond to the different values of $K$ as shown in the legend of panel (b).
    }
\end{figure}

Figure~\ref{fig:pt-tebd-results-a} shows the dynamics of a spin chain of length $N=7$ for $K=0, \ldots, 3$ attached environments.
We can observe in Fig.~\ref{fig:pt-tebd-results-a} that the spreading of the initial excitation is slowed down with every additional environment coupled to the chain.
Although the PT-TEBD algorithm does not allow for direct interaction among the environments, it still captures the correlations among the environments that build up through the chain.
These correlations are reflected in the bond dimension of the augmented MPS.
Figure~\ref{fig:pt-tebd-results-b} shows the maximal bond dimension of each bond, for different $K$.
While we can observe significant growth of bond dimensions with a growing number of environments $K$, their values are still well below the possible maximum that is given by the system Hilbert space dimension ($d=2$) and the bond dimensions of the PT-MPOs ($\chi=11$).
For $K=3$ this would give a maximum at bond 2 with $(\xi d^2)^2=1936$ which is an order of magnitude larger than the observed value of 167.

We note that a few alternative methods for chains of non-Markovian open quantum systems exist in the literature~\cite{Makri2018a, Makri2018, Suzuki2019, Purkayastha2020, Makri2021, Kundu2021, Kundu2021a, Flannigan2021, Bose2021, kundu_pathsum_2023}.
While a quantitative comparison with these methods would need to be done case by case, qualitatively the PT-TEBD approach stands out in two ways.
First, it approaches the challenging many-body problem sequentially by systematically reducing the numerical complexity originating from system-environment correlations through compression of the PT-MPO, before integrating them into the full many-body problem.
Second, the method is agnostic about the origin of the process tensor, i.e. it is applicable to any environment, given its PT-MPO.
The PT-TEBD method has been applied successfully to a 21 site long XYZ-Heisenberg spin chain with strongly coupled bosonic environment on every site using only moderate computational resources, and, it has been proven useful for studying thermalization in a strongly coupled open many-body quantum system~\cite{fux_tensor_2023}.

\subsection{Mean-field open quantum systems}
\label{sub:mean-field}
\begin{figure}
	\phantomsubfloat{\label{fig:mean-field-schematic-a}}%
	\phantomsubfloat{\label{fig:mean-field-schematic-b}}%
	\phantomsubfloat{\label{fig:mean-field-schematic-c}}%
	\includegraphics[width=0.48\textwidth]{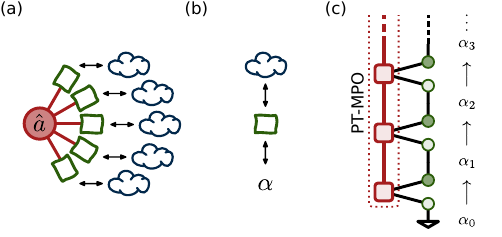}
	\caption{\label{fig:mean-field-schematic}
		(a)~Schematic for the many-to-one network of a central boson model: each emitter (green squares) couples to a single bosonic mode \(\hat{a}\) (red circle) as well as a harmonic environment (blue clouds).
		(b)~Mean-field theory reduction to a single emitter interacting with a classical field \(\alpha= \langle \hat{a}^{\vphantom{\dagger}}_{} \rangle\).
		(c)~Tensor network for the PT-MPO method with concurrent dynamics of the cavity field.
		The field \(\alpha_i\) at \(t=t_i\) is used in Eq.~\eqref{eq:mean-field_H} to construct the system propagators at that time.
		Once the molecular system has been evolved to \(t=t_{i+1}\), the resulting expectation \(\langle \hat{\sigma}^-_{} \rangle_{i+i}\) is used to integrate the field \(\alpha_i\to\alpha_{i+1}\)	according to Eq.~\eqref{eq:mean-field_a}.
	}
\end{figure}

A second type of many-body open quantum system that can be addressed in \oqupy{} are models with many-to-one or star-like topologies, see Fig.~\ref{fig:mean-field-schematic-a}.
We discuss this scenario by an example of a central boson model as was recently used to study the non-Markovian dynamics of organic polaritons~\cite{fowler2022}.
Here, a large number \(N\) of organic emitters couple to independent structured environments as well as a common photon mode (with bosonic operator \(\hat{a}^{\vphantom{\dagger}}_{}\)).
The emitters are modeled as two-level systems (Pauli matrices \(\hat{\sigma}^\alpha_n\)) and the light-matter dynamics is governed by 
\begin{align}
		\hat{H}_\mathrm{S}= \omega_c \hat{a}^{\dagger}_{}\hat{a}^{\vphantom{\dagger}}_{} &+ \sum_{n=1}^N
		\left[\frac{\omega_0}{2} \hat{\sigma}^z_n +  \frac{\Omega}{2\sqrt{N}} \left(
		\hat{a}^{\vphantom{\dagger}}_{} \hat{\sigma}^+_n + \hat{a}^{\dagger}_{} \hat{\sigma}^-_n
		\right)\right],
\end{align}
where \(\omega_c\) and \(\omega_0\) are the cavity and emitter frequencies, and \(\Omega\) is the collective light-matter coupling strength.
Each emitter is also coupled to a harmonic bath,
\begin{align}
		\hat{H}^{(n)}_\mathrm{IE} = \sum_{j} 	\left[	\nu_{j} \hat{b}^{\dagger}_{j,n}
		\hat{b}^{\vphantom{\dagger}}_{j,n} + \frac{\xi_{j}}{2}
		\left(\hat{b}^{\vphantom{\dagger}}_{j,n}+\hat{b}^{\dagger}_{j,n}\right)\hat{\sigma}^z_n\right].
\end{align}
This captures the local vibrational environment of organic molecules, which may typically be a low frequency continuum of modes described by, for example, an Ohmic spectral density \(J(\nu)\).
In addition, there are cavity losses with rate \(\kappa\) and individual pumping \(\Gamma_\uparrow\) and dissipation \(\Gamma_\downarrow\) of the emitters; these can be included as incoherent processes under the Markovian approximation. 

In order to solve the many-molecule-cavity dynamics, we use mean-field theory to reduce the dimension of the problem.
In this approach, one neglects correlations in the many-body state, i.e. one assumes a product
\begin{align}
		\rho = \rho_a \otimes \bigotimes_{n=1}^N \rho_n,
\end{align}
where \(\rho_a\) and \(\rho_n\) are reduced density matrices for the cavity and \(n^{\text{th}}\) emitter, respectively.
This reduces the problem to the coupled dynamics of the molecular mean-field Hamiltonian~\cite{fowler2022} 
\begin{align}
	\hat{H}_\mathrm{MF} = \frac{\omega_0}{2} \hat{\sigma}^z +
	\frac{\Omega}{2\sqrt{N}}(\langle \hat{a}^{\vphantom{\dagger}}_{} \rangle 
	\hat{\sigma}^+ 	+ \langle \hat{a}^{\vphantom{\dagger}}_{} \rangle^* \hat{\sigma}^-), 
	\label{eq:mean-field_H}
\end{align}
combined with evolution of the field expectation
\begin{align}
	\partial_t \langle \hat{a}^{\vphantom{\dagger}}_{} \rangle
	&= -(i\omega_c+\kappa) \langle \hat{a}^{\vphantom{\dagger}}_{}\rangle
	-i\frac{\Omega\sqrt{N}}{2} \langle \hat{\sigma}^- \rangle \text{.}
	\label{eq:mean-field_a} 
\end{align}
Here \(\langle \hat{\sigma}^- \rangle=\langle \hat{\sigma}^-_{n} \rangle\) is the average of any of the identical spins. 

Thus, by propagating a \emph{single} spin with \(\hat{H}_\mathrm{MF}\) subject to the vibrational environment and individual pump and dissipation described above, one can effectively simulate the \(N\)-molecule system using the TEMPO or PT-TEMPO method provided that at each time step the field \(\langle a^{\vphantom{\dagger}}\rangle\) is evolved according to Eq.~\eqref{eq:mean-field_a}, as schematically depicted in Fig.~\ref{fig:mean-field-schematic-c}.

\begin{figure}
    \phantomsubfloat{\label{fig:mean-field-results-a}}%
    \phantomsubfloat{\label{fig:mean-field-results-b}}%
    \begin{tikzpicture}[every node/.style={font=\sffamily}]
    \node (a) at (0,0) {\includegraphics[width=0.495\textwidth]{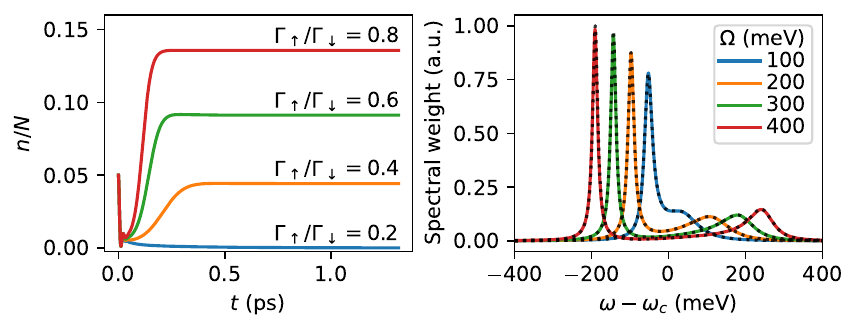}};
    \node [shift={(1.48,-.53)}] at (a.north west) {(a)};
    \node [shift={(5.74,-.53)}] at (a.north west) {(b)};
    \end{tikzpicture}
    \caption{\label{fig:mean-field-results}
        (a)~Scaled photon number \(n/N=|\langle	a\rangle|^2/N\) below (\(\Gamma_\uparrow = 0.2 \Gamma_\downarrow\)) and above (\(\Gamma_\uparrow \geq 0.4\,\Gamma_\downarrow\)) a lasing transition at \(\Omega=200\)\,meV, \(T=300\)~K and \(\Delta = \omega_c-\omega_0=-20\)\,meV~\cite{fowler2022}.
        Dynamics were calculated from an initial state with the spin down and a small number of photons \(n_0/N=0.05\). 
        (b)~Spectral weight (absorption) when \(\Gamma_\uparrow=0\).
        At each light-matter coupling, an analytic result~\cite{kirton2015} is shown as a dotted line, and the result~\cite{fowler2022} of a mean-field PT-TEMPO calculation as a solid line.
        The PT-TEMPO numerical parameters used were
        $\code{tcut}=0.1\,\ps$,
        $\code{dt}=4.0\times 10^{-4}\,\ps$,
        $\code{epsrel}=5.0\times10^{-12}$.
    }
\end{figure}

Figure~\ref{fig:mean-field-results-a} illustrates this method in calculating photon number dynamics for different pump ratios \(\Gamma_\uparrow/\Gamma_\downarrow\).
This can be used, for example, to determine the dependence on the threshold for organic lasing on cavity detuning \(\omega_c-\omega_0\) and light-matter coupling strength \(\Omega\) (see Ref.~\onlinecite{fowler2022} for details).

To verify the numerics, we calculated the absorption spectrum (see Sec.~\ref{sub:multi-time-correlations}) for the system without pumping (\(\Gamma_\uparrow=0\)), for which an analytical result is known~\cite{kirton2015}.
Figure~\ref{fig:mean-field-results-b} shows excellent agreement between this result and that of a PT-TEMPO with mean-field calculation.

\begin{figure}
    \centering
    \includegraphics[width=0.35\textwidth]{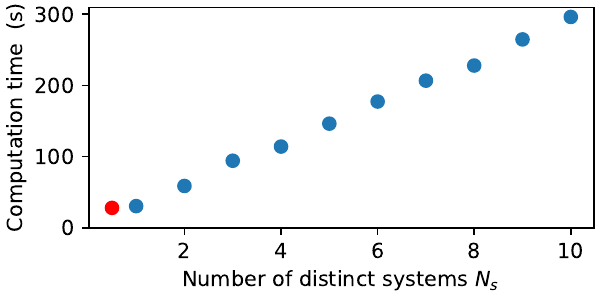}
    \caption{\label{fig:mean-field_scaling} 
        Scaling of mean-field dynamics computation with number of distinct systems (emitters).
        A comparable calculation without mean-field integration is included as a red point.
	}
\end{figure}

Note that in the above we took the emitters to be identical.
The mean-field approach is, however, not limited to this case and so can describe e.g. multiple molecular species in a single cavity~\cite{georgiou2021}.
The cost is a separate PT-TEMPO (or TEMPO) computation for each distinct \(\rho_n\).
Figure~\ref{fig:mean-field_scaling} shows that the mean-field integration adds negligible overhead to a standard PT-TEMPO computation, whilst the total computation time for \(N_s\) types of emitters scales linearly with \(N_s\).

Mean-field dynamics is accessible in \oqupy{} from an instance of the \code{MeanFieldSystem} class.
This comprises one or more \code{TimeDependentSystemWithField} objects and a callable \code{field\_eom}.
The former describes systems (emitters) with different mean-field Hamiltonians (such as Eq.~\eqref{eq:mean-field_H}), and the latter specifies the field equation of motion (e.g. Eq.~\eqref{eq:mean-field_a}).

In this section we discussed a central boson model, for which the mean-field approximation can be shown~\cite{carollo2021} to be exact as \(N\to\infty\).
The same approach can be applied in other situations where mean-field theory is known to either be exact, or become a good approximation, in this limit.
This includes central spin models~\cite{makri1999} and models with all-to-all interactions~\cite{fiorelli2023} that arise, for example, in the context of waveguide and cavity-QED with cold atoms~\cite{lewenstein2007, chang2018}.


\section{Other related methods}
\label{sec:other-methods}
The methods and use cases presented in the previous Section~\ref{sec:use-cases}, all start from a pre-computed PT-MPO and are thus applicable to any type of environment for which such an PT-MPO can be obtained, regardless of the nature of the environment or the method with which the PT-MPO has been computed.
In this section we give a brief introduction to further useful methods that are available in \oqupy{}, but are more specific as they require that the environment is a linearly coupled Gaussian bosonic bath.

\subsection{Environment dynamics}
\label{sub:environment-dynamics}
A key step in PT construction, and many approaches to simulating open quantum systems, is tracing over the environment degrees of freedom.
However, for non-Markovian processes these degrees of freedom by definition play a non-trivial role.
In tracing them out we lose key insight into the interplay between system and environment.
This could involve, for example, any engineered effect of the system on the environment such as with quantum thermal machines~\cite{Mitchison2019, Binder2019}---where distinct thermodynamic effects beyond weak coupling can be observed~\cite{Brenes2019, strasberg2016nonequilibrium}---or the formation of system-environment bound states such as polarons~\cite{Mahan2000, silbey1984variational, harris1985variational, devreese2009frohlich}.

The loss of access to environment information has previously put process tensor approaches at a disadvantage to methods such as the reaction coordinate or chain mappings~\cite{Iles-Smith2014, Chin2010, Prior2010} which capture non-Markovian effects by augmenting the system with certain environmental degrees of freedom; here direct insight into environment dynamics can be gained.
However, for the widely applicable case of linearly coupled bosons in a Gaussian initial state the bath correlations can be directly calculated from system correlation functions via an integral transform~\cite{Gribben2021}.
This relation is general and can be used with any technique capable of generating system correlations.
The PT-MPO approach is naturally placed to efficiently compute the many-system correlations necessary for the calculation, as outlined in Section~\ref{sub:multi-time-correlations}.

To demonstrate this approach we shall now compute the change in energy of the environment resolved by mode frequency for the biased spin-boson model with an Ohmic environment.
The total Hamiltonian is thus again of the form of Eq~\eqref{eq:tempo-hamiltonian}, with the system Hamiltonian $\hat{H}_\mathrm{S} = \frac{\epsilon}{2} \hat{\sigma}_z + \frac{\Omega}{2} \hat{\sigma}_x$.
This could describe, for example, a semiconductor quantum dot driven by a laser with strength $\Omega=1.0\,\ips$ and bias  $\epsilon=2.0\,\ips$ coupled to the phonon modes in the medium.
The phonons are described by a bath at temperature $T=1.0\,\ips \,(=7.64\,\mathrm{K})$ with a spectral density $J(\omega)$ given by Eq.~\eqref{eq:spectral-density-power-law}, where $\zeta=1.0$, $\alpha=0.05$, and $\omega_c=10.0\,\ips$.
The initial state of the system is $\rho^\mathrm{S}_0 = \ketbra{\downarrow}{\downarrow}$.

\begin{figure}
    \centering
	\phantomsubfloat{\label{fig:bath-dynamics-results-a}}%
	\phantomsubfloat{\label{fig:bath-dynamics-results-b}}%
	\includegraphics[width=0.49\textwidth]{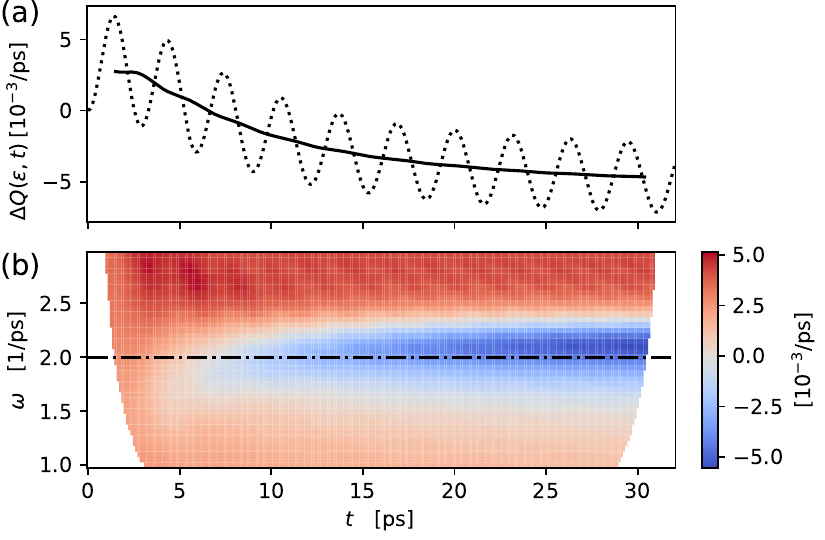}
    \caption{\label{fig:bath-dynamics-results}%
    	Change of energy in environment modes after the quench (by coupling system and environment) at $t=0$.
    	(a)~The change of energy in the mode interval $[\epsilon-\delta/2, \epsilon+\delta/2]$.
    	The dotted line shows the raw result, while the solid line shows the period-averaged result.
    	(b)~The period-averaged change of energy in modes, $\Delta Q(\omega,t)$, as defined in Eq.~\eqref{eq:heat_delta}. The lack of data at early and late times is a result of the period-averaging.
	}
\end{figure}

The environment is taken to be a continuum such that the coupling between the system and any single mode is infinitesimal.
As such it makes no sense to investigate the change in energy of single modes.
Instead we compute the change in energy over a range of modes with bandwidth $\delta$.
This is given by
\begin{equation}
	\label{eq:heat_delta}
	\Delta Q(\omega,t) = \sum_k\int_{\omega-\delta/2}^{\omega+\delta/2} \dd\nu\, \delta(\nu-\omega_k) \omega_k \Delta n_k(t),
\end{equation}
where $\Delta n_k(t)=\langle \hat{b}_k^\dagger(t)\hat{b}_k(t)\rangle-\langle \hat{b}_k^\dagger(0)\hat{b}_k(0)\rangle$ is the change in occupation of environment mode $k$ from the initial state to time $t$; for the results presented here we fix $\delta=0.1\,\ips$.
We plot this as a function of $t$ and $\omega$ in Fig.~\ref{fig:bath-dynamics-results-b} and see that most bath modes show a modest increase in energy, this is associated with the heating that arises from the interaction being quenched at $t=0$.
A notable departure from this trend occurs in the vicinity of modes resonant to the laser bias frequency $\epsilon$ which instead, after an initial heating, show a reduction in energy.
This can be understood as the bath providing the additional energy lacking in the laser drive for the transition to be driven resonantly.
This is highlighted in the upper panel where we focus on the change in energy in the vicinity of modes resonant with $\epsilon$.
Also visible here are long-lived oscillations in the heat exchanged, this feature is present in all frequency bands and is a result of the relaxation time of these bands increasing as the width of the band is reduced.

In \oqupy{} two-time bath correlation functions can be calculated using the \code{bath\_dynamics} module. Currently this module consists of a single class, \code{TwoTimeBathCorrelations}, which has the capability of calculating any correlation function of the form
\begin{equation}
	\langle \hat{b}_{k_2}^{(\dagger)} (t_2) \hat{b}_{k_1}^{(\dagger)} (t_1) \rangle,
\end{equation}
where $t_2 \geq t_1$.
All that is required is a \code{BaseSystem}, a \code{Bath}, and a \code{BaseProcessTensor}.
The necessary system correlations, if not provided, are determined and computed automatically.
A single set of these system correlations can be used to calculate a variety of bath correlation functions by simply adjusting the integral transform applied.
By defining the two-time bath correlations as a class, any system correlations computed for a given bath correlation can be stored within the class object and re-used in subsequent bath correlation calculations for a much faster computation.

Here we have highlighted how \oqupy{} can be used to calculate two-time bath correlation functions.
In contrast to the other use cases, this relies a on derived relation assuming a specific form of interaction and bath; namely a linear coupling to a Gaussian bath of bosons.
Although restrictive, this form of environment is ubiquitous in literature in the form of spin-boson models~\cite{Leggett1987,weiss2012quantum}.
In future it would be interesting to expand the \code{bath\_dynamics} class to handle more general bath correlations. 
For example, computation of out-of-time-order correlators~\cite{Xu2024}, in particular one mixed between system and bath, could give a deeper insight into how information on the system is scrambled by a non-Markovian environment.
In Ref.~\onlinecite{Gribben2021} a recipe is provided that gives the transform of system correlations necessary to compute this along with any other bath correlations desired.

\subsection{Time evolving matrix product operator method}
\label{sub:tempo}

\hide{
\begin{figure}
    \includegraphics[width=0.49\textwidth]{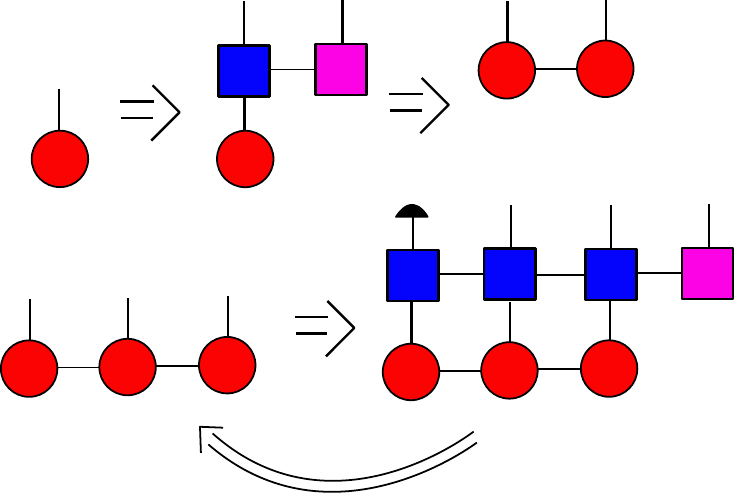}
    \caption{\label{fig:tempo-network} \sketch{Sketched diagram, needs formatting.}
    Diagrammatic approach of the contraction method used in the TEMPO algorithm.
    Starting with he initial state (red circle), this is contracted with the first influence functional (blue square) and system propagator (pink square) as shown.
    This gives an MPS state in time, where every step to the left indicates one step back in time.
    This process is repeated until the memory cutoff is reached, then a trace cap is applied to the final influence functional, maintaining the size of the MPS state.}
\end{figure}
}

In previous sections we have seen how process tensors can be useful for studying many situations which commonly arise when simulating open quantum systems.
An alternative approach for the case of Gaussian bosonic environments is to contract the tensor network shown in Fig.~\ref{fig:pt-tempo-network-a} in a different way to directly obtain the system dynamics instead of creating a PT-MPO.
This is the original time evolving matrix product operator method (TEMPO) algorithm~\cite{Strathearn2018} as mentioned in Sec.~\ref{sub:pt-tempo}. 
This loses the advantages of the process tensor approach for being able to efficiently solve many problems which are related by having the same bath Hamiltonian.
It can, however, be useful in cases where only a single simulation is required.

The key difference between TEMPO and PT-TEMPO is the order of contractions of the tensor network.
In TEMPO we contract the tensor network row by row, giving an MPS after each contraction step.
This MPS then gives access to the state of the system which can be recorded to calculate dynamics.
As with PT-TEMPO, a memory cut-off can be used to limit how many history points are tracked.
This means the MPS will grow in size until it reaches the cut-off where it will reach its maximum size.
After this growth phase, a propagator can be defined as all subsequent layers are identical (for time independent Hamiltonians).
After application of the propagator the MPS will grow one extra site to the right, and we contract the leftmost tensor to maintain the size of the MPS.
This process can then be repeated until the desired time step is reached.

The TEMPO method can be accessed in \oqupy{} through the \code{tempo\_compute()} function.
This function takes a \code{Bath} (with $J(\omega)$, $T$, and $\hat{S}$) and \code{System} or \code{TimeDependentSystem} object (with $\hat{H}_\mathrm{S}$) and returns a \code{Dynamics} object which encodes the evolution of the reduced system density matrix. 

\subsection{Gibbs TEMPO}
\label{sub:gibbs-tempo}
Although the functionality of \oqupy{} is primarily concerned with non-equilibrium systems, the tensor network methods that have been introduced can also be used to calculate equilibrium properties of open quantum systems. 
Specifically, if a system and Gaussian environment are collectively in a Gibbs state, i.e in thermal equilibrium at some temperature $T=1/\beta$, the TEMPO method discussed in the previous section can be used, with minor alterations, to calculate the reduced Gibbs state of the system (without resorting to weak coupling, or separability assumptions).
The Gibbs TEMPO method~\cite{Chiu2021} achieves this by performing an ``evolution'' along the imaginary time axis from time $t=0$ to time $t = i \beta$.

The main difference between Gibbs TEMPO and standard TEMPO is that here the influence functional is written in terms of operators rather than superoperators, such that the legs of the influence tensors have dimension $d$, equal to that of the dimension of the system Hilbert space, rather than $d^2$.
If the goal is to simply calculate a thermal steady state of an open quantum system, this makes Gibbs TEMPO the more efficient option over standard TEMPO. 
Another difference is that the environment correlation functions do not decay on the imaginary time axis and hence no memory cutoff $\code{tcut}$ can be used.
Also, there is no initial state required as an input to Gibbs TEMPO.

In \oqupy{} Gibbs TEMPO is accessed through the \code{gibbs\_tempo\_compute()} function, which takes
a \code{System}, a \code{Bath}, and a \code{GibbsParameters} object, and returns the reduced Gibbs state in the form of a $d \times d$ array.
The \code{GibbsParameters} class takes the number of imaginary time steps \code{num\_steps} and a relative singular value cut-off threshold \code{epsrel}.
The \code{Bath} object must have been initialized using a spectral density (i.e. either a \code{CustomSD} or \code{PowerLawSD} object).


\section{Future directions}
\label{sec:future-directions}

The versatility of version 0.5 of \oqupy{} is the result of continuous extension of its functionality in the past couple of years.
We intend to continue active development and maintenance of this open source project, and for this also welcome contributions from outside the current collaborations.
We hence close in this section with a brief discussion of potential further extensions of \oqupy{} that we believe could be useful for research in fields related to non-Markovian open quantum systems.

As discussed above, the process tensor formalism is not restricted to a particular form of the environment.
To fully leverage this fact in \oqupy{}, a useful extension would thus be to implement methods that yield PT-MPOs for environments other than Gaussian bosonic environments~\citePTcreation{}.
A particularly versatile method would be the \emph{automated compression of environments} (ACE) method~\cite{Cygorek2021}, which allows an efficient computation of PT-MPOs for a range of different environments, including anharmonic non-Gaussian bosonic, fermionic, and spin environments.

Beside additional methods for the creation of PT-MPOs, there are a series of methods for manipulating existing PT-MPOs that would be of practical value to develop, such as methods to combine, cut, and course grain PT-MPOs.
This would grant further flexibility to the construction and integration of process tensors in use cases such as above.
Also, to facilitate the modeling of realistic environments, which may be highly structured, one can consider coarse-graining schemes~\cite{Lorenzoni2024} for spectral densities.
In \oqupy{} these could be implemented as an interface in which one inputs a complex spectral density---derived from experimental data or otherwise---and an effective, simplified spectral density is used to construct a corresponding \texttt{Correlations} object that is numerically tractable.

Another set of tools that would greatly enrich the functionality of \oqupy{} are methods that create and employ time translational invariant PT-MPOs, which encode an PT-MPO of arbitrary length in a single time translational unit cell~\cite{cygorek2023, link2023}, in analogy to infinite MPS in space~\cite{Vidal2007a}.
Such time translational invariant process tensors could be particularly useful for the direct study of non-equilibrium steady states of non-Markovian open quantum systems, with possibly interesting applications in the field of quantum thermodynamics.

Finally, we emphasize that the methods implemented in \oqupy{} and described in this article have been developed and employed in first studies only recently~\citeallmethods, with great scope for their application in many other problems that have previously been inaccessible.
The primary purpose of \oqupy{} thus is, and continues to be, the low-barrier access to those numerical methods and to foster research in fields such as quantum chemistry, quantum technology, and quantum thermodynamics.


\subsubsection*{Acknowledgments}
We thank the Unitary Fund community for their support.
GEF acknowledges support from EPSRC (EP/L015110/1) and from ERC under grant agreement n.101053159 (RAVE).
JB acknowledges support from the Laidlaw Foundation (Leadership and Research Programme scholarship).
EPB acknowledges support from the Irish Research Council (GOIPG/2019/1871).
DG acknowledges support from the QuantERA II Programme that has received funding from the European Union’s Horizon 2020 research and innovation programme under grant agreement n.101017733 (“QuSiED”).
PRE acknowledges support from Science Foundation Ireland (21-FFP-10142).
EDCL acknowledges support from EPSRC (EP/T517938/1).
RdW acknowledges support from EPSRC (EP/W524505/1).
BWL and JK acknowledge support from EPSRC (EP/T014032/1).

\subsubsection*{Author contributions}
The software project and writing of this article has been led by GEF and co-led by PFW.
The TEMPO and PT-TEMPO method has been implemented by GEF and extended by EDCL and PK, building on precursor code by AS, DK, and PK.
Multi-time correlations functionality has been contributed by RdW.
The gradient functionality has been contributed by EPB, PRE, and EON.
Multiple environment and environment dynamics functionality has been contributed by DG.
The PT-TEBD method has been implemented by GEF, building on precursor code by DK.
Mean-field functionality has been implemented by PFW, and extended by JB.
Gibbs TEMPO has been implemented by AS.
BWL and JK supervised and coordinated the scientific development.
All code contributors have written the corresponding sections in the article.

\subsection*{Data availability}
The data that support the findings of this study are openly
available on Zenodo at \href{https://doi.org/10.5281/zenodo.4428316}{https://doi.org/10.5281/zenodo.4428316}, reference
number 7243607.


%

\end{document}